\documentclass{jfm}
\usepackage{graphicx}

\usepackage{siunitx} 
\usepackage{amsmath} 
\usepackage{longtable} 
\usepackage{multirow} 
\usepackage[caption=false]{subfig}
\usepackage{wasysym} 
\usepackage{pdflscape} 
\usepackage{color}

\usepackage{tikz}
\usepackage{pgfplots}
\pgfplotsset{compat=1.8}

\newcommand{\imag}{\mathrm{i}} 
\newcommand{\expo}{\mathrm{e}} 
\renewcommand{\vec}[1]{\boldsymbol{#1}} 
\newcommand{\tens}[1]{\mathsfbi{#1}} 
\newcommand\Str{\mbox{\textit{St}}}  

\shorttitle{Receptivity of the turbulent precessing vortex core}
\shortauthor{J. S. M\"{u}ller, F. L\"{u}ckoff, P. Paredes, V. Theofilis and K. Oberleithner}

\title{Receptivity of the turbulent precessing vortex core: synchronization experiments and global adjoint linear stability analysis}

\author{J. S. M\"{u}ller\aff{1}
  \corresp{\email{jens.mueller@tu-berlin.de}},
  F. L\"{u}ckoff\aff{1},
  P. Paredes\aff{2},
  V. Theofilis\aff{3},
 \and K.~Oberleithner\aff{1}}

\affiliation{\aff{1}Laboratory for Flow Instabilities and Dynamics, Technische Universit\"{a}t Berlin, M\"{u}ller-Breslau-Str.~8, 10623 Berlin, Germany
\aff{2}National Institute of Aerospace, 1100 Exploration Way, Hampton, Virginia 23666, USA
\aff{3}School of Engineering, University of Liverpool, The Quadrangle, Brownlow Hill, Liverpool L69 3GH, UK}

\begin{document}

\maketitle

\begin{abstract}

The Precessing Vortex Core (PVC) is a coherent structure that can arise in swirling jets from a global instability. In this work, the PVC is investigated under highly turbulent conditions. The goal is to characterize the receptivity of the PVC to active flow control, both theoretically and experimentally.
Based on stereoscopic particle image velocimetry and surface pressure measurements, the experimental studies are facilitated by Fourier decomposition and proper orthogonal decomposition. The frequency and the mode shape of the PVC are extracted and a very good agreement with the theoretical prediction by global linear stability analysis (LSA) is found. By employing an adjoint LSA, it is found that the PVC is particularly receptive inside the duct upstream of the swirling jet. Open-loop zero net mass flux actuation is applied at different axial positions inside the duct with the goal of frequency synchronization of the PVC. The actuation is shown to have the strongest effect close to the exit of the duct. There, frequency synchronization is reached primarily through direct mode-to-mode interaction. Applying the actuation farther upstream, synchronization is only achieved by a modification of the mean flow that manipulates the swirl number. These experimental observations match qualitatively well with the theoretical receptivity derived from adjoint LSA.
Although the process of synchronization is very complex, it is concluded that adjoint LSA based on mean field theory sufficiently predicts regions of high and low receptivity. Furthermore, the adjoint framework promises to be a valuable tool for finding ideal locations for flow control applications.
\end{abstract}

\begin{keywords}
Authors should not enter keywords on the manuscript, as these must be chosen by the author during the online submission process and will then be added during the typesetting process (see http://journals.cambridge.org/data/\linebreak[3]relatedlink/jfm-\linebreak[3]keywords.pdf for the full list)
\end{keywords}

\section{Introduction}
Turbulent swirling flows undergoing vortex breakdown have a wide range of application in engineering flows. Amongst others, they occur on delta wings at high angle of attack \citep{Peckham1957}, they are exploited for aerodynamically stabilizing the reaction zone of lean premixed flames in combustion chambers of modern gas turbines \citep{Syred1974}, or they emerge in hydro turbines at off-design conditions \citep{Nishi1980}. 

Vortex breakdown is often accompanied by a strong coherent structure which is known as the Precessing Vortex Core (PVC). It is characterized by a single-helical vortex that winds around the vortex breakdown bubble in downstream direction, thereby oscillating at a well-defined precession frequency.

The PVC may be a desired structure or it may be detrimental to the purpose of the application. In combustion systems, the PVC has been investigated in numerous studies. The PVC may alter the flame stability and dynamics \citep{Stoehr2012a,Oberleithner2014,Stoehr2017}, it may couple with thermoacoustic instabilities \citep{Moeck2012,Terhaar2016,Ghani2016} or it may influence the mixing of fuel and air \citep{Stoehr2015,Terhaar2015}. So far, only qualitative trends on these different phenomena could be determined and the quantitative impact of the PVC is still unclear. The question remains whether the PVC can be used in a beneficial way to reduce pollutant emissions or increase flame stability. Hydro power is another important application where the PVC plays a crucial role. It occurs in the outflow of Francis turbines running at part load. Several studies confirm that the PVC, also known as the part-load vortex rope, may couple with the hydroacoustics of the pipe system and induce severe pressure pulsations that hinders the safe operation of the hydro power plant \citep{Pasche2017}. This results in low turbine flexibility and small operational range, severely affecting the usability of hydro turbines which are increasingly used to balance the transient natural fluctuations of renewable energy sources such as wind and solar power (see e.g. \cite{Doerfler2012}). Hence, from an application point of view, the PVC is an important fluid-dynamical phenomenon that needs to be controlled in an effective manner. For this purpose it is vital to understand how the PVC is generated, what the mechanisms of manipulation are, and how the control is most effectively done.

In the recent years the fundamental understanding of the origin of the PVC has substantially increased. Based on experimental studies, \cite{Liang2005} were among the first to demonstrate that the PVC is the manifestation of a global hydrodynamic instability that occurs from a supercritical Hopf bifurcation. They further suggested that this global mode is triggered by a region of absolute instability enclosed in the vortex breakdown bubble. Further evidence of this hypothesis was provided by the direct numerical simulations of laminar vortex breakdown conducted by \cite{Ruith2003} and the subsequent local spatio-temporal stability analysis conducted by \cite{Gallaire2006}. An extension to turbulent vortex breakdown was provided by \cite{Oberleithner2011}, where the stability analysis was applied to the turbulent mean flow. \cite{Rukes2015a} extended this study to a wider swirl number range. Despite some differences regarding the detailed mechanisms driving the PVC, all these studies indicate that the recirculation bubble induced by vortex breakdown at strong swirl levels is the main requirement for the formation of the PVC that is directly linked to absolute and global instability.  

In recent years, the local and global LSA performed on the time-averaged mean flow turned out to be a very useful approach to study the formation of the PVC in turbulent flows. Within the framework of the \textit{local} LSA, \cite{Chomaz2005} derived a frequency selection criterion for linear global modes in spatially developing flows, under the assumption of a parallel or weakly non-parallel flow. This criterion was extended to nonlinear steep global modes by \cite{Pier2001}. Considering a laminar baseflow solution of the axisymmetric vortex breakdown bubble, \cite{Gallaire2006} showed that the nonlinear criterion enables to reproduce the oscillation frequency of a PVC determined from a direct numerical simulation. They concluded that the PVC spiral instability is the manifestation of a nonlinear steep global mode that arises in the wake of the breakdown bubble. \cite{Qadri2013} computed the structural sensitivity of the same baseflow based on a global framework and concluded that the linear global mode that is selected upstream of the breakdown bubble is more influential. \cite{Oberleithner2011} utilized the local LSA on experimental data of a turbulent swirling jet and found a good agreement between the spatial modes obtained by local LSA and the modes obtained by Proper Orthogonal Decomposition (POD). In follow-up studies related to turbulent flows, it was shown that an eddy viscosity closure is crucial to correctly predict the PVC frequency and mode structure \citep{Oberleithner2014,Rukes2016}. To clarify the discussion about the selection mechanism that applies to the PVC, \cite{Rukes2015a} conducted transient experiments and tracked the wavemaker location as a function of swirl number. In agreement with \cite{Qadri2013}, they determined the PVC to be selected at a position shortly upstream of the breakdown bubble. Abandoning the parallel flow assumption, which is not rigorously justified for swirling flows, the \textit{global} LSA requires much higher computational cost. As these demands are increasingly met, global LSA has found a wider application in the last decades \citep{Theofilis2011}. Based on experimental data, \cite{Paredes2016} were able to reproduce the global mode shapes downstream of the mixing section of a combustor with global LSA. \cite{Tammisola2016} as well as \cite{Kaiser2018} followed in demonstrating the applicability and validity of global LSA on highly turbulent flows with a dominating PVC instability. Overall, the excellent match between global mean flow eigenmodes and the PVC structure clearly confirms that the PVC is indeed the manifestation of a global mode, which can be very well determined from global and local stability analysis based on the mean of the fully turbulent flow. Inconsistency remains with regard to the mechanisms that generate the PVC, which is either related to nonlinear mechanisms in the wake of the vortex breakdown bubble or to linear mechanisms upstream of the bubble.

Within the framework of the global LSA, the adjoint mean flow and the adjoint modes have become the focus of interest in understanding the driving mechanisms and guiding future control strategies. They have been used by \cite{Giannetti2007}, \cite{Marquet2008} and \cite{Meliga2012,Meliga2016} in the context of wake flow vortex shedding of different cylinder geometries to investigate the impact of passive flow control devices on the frequency of the shedding or its suppression. They validated their results using experimental data from \cite{Strykowski1990}, \cite{Parezanovic2012} and \cite{Meliga2016} and showed that primarily the adjoint mean flow is contemplable for estimating the impact of these devices. Higher values in the adjoint corresponded to greater changes in the modal structure when a very small solid body was placed at this particular location. The results imply that the impact of a passive control device essentially acts through a steady forcing of the mean flow which in turn changes the modes. Specifically considering active control measures, \cite{Magri2014} used the adjoint modes to investigate a simple numerical thermoacoustic system comprising a diffusion flame to find regions of the flame which are most receptive to open-loop, i.e. unsteady and periodic, forcing. Regarding the PVC, \cite{Qadri2013} considered the laminar base flow solution of vortex breakdown and determined the maximum structural sensitivity to be located between the domain inlet and the breakdown bubble. The same configuration was later used by \cite{Pasche2018} to develop predictive control based on the adjoint of the global mode. They found that axial air injection along the jet centerline may effectively suppress the PVC, which is in line with experiments conducted in swirl combustors \citep{Terhaar2013b} and hydro turbines \citep{Kirschner2010}.

An open question with regard to the modeling of the PVC concerns the correct treatment of the domain boundary upstream of the vortex breakdown. Some works demonstrate that this region is crucial for the formation of the PVC. \cite{Tammisola2016} conducted direct numerical simulations of a fuel injector system and determined the PVC to be also located inside the fuel injector. Moreover, they showed that the receptivity to open-loop forcing reaches far upstream of the injector. \cite{Kaiser2018} calculated the adjoint modes of the PVC in a more complex swirl injector geometry based on large eddy simulations. Within their study, regions of high receptivity were also found inside the injector. Overall, the adjoint modes reveal the importance of the upstream region for the generation of the PVC and, additionally, identify the best location for its control. Inconsistency between different studies related to the PVC formation may, therefore, stem from the different treatment of the upstream boundary conditions or from different placement of the domain boundaries.

Due to the high relevance of the PVC for technical applications, several control methods have been employed. Within the scope of hydro power, passive control methods succeeded to suppress the instability by reducing the swirl intensity but they are only applicable for a very narrow operational regime and create additional undesirable hydraulic losses \citep{Kurokawa2010}. The most effective active control solution uses constant water or air injection to suppress the PVC. Experiments however show that the suppression requires more than $10\%$ of the total main flow discharge, which implies an unacceptably large loss of the turbine efficiency \citep{Kirschner2010}. For combustion applications, the PVC may be controlled via axial air injection \citep{Terhaar2015}, acoustic forcing \citep{Terhaar2016} or through modification of the flame shape \citep{Oberleithner2015}. In all cases, suppression of the PVC was achieved through modification of the mean flow field, as it was \textit{a posteriori} justified by LSA. Another recent control approach is tailored to suppress the instability by phase-opposition control \citep{Kuhn2016}. Active control was exerted by a thin actuator lance placed at the jet centerline. The actuator was traversed in axial direction and the receptivity of the PVC to the forcing was determined from lock-in experiments. The study indicated the highest receptivity upstream of the vortex breakdown bubble, but ambiguities remained due to the passive influence of the actuator on the PVC frequency. In a consecutive study, the same control method was employed to control the PVC in a swirl combustor \citep{Lueckoff2018,Lueckoff2019}. It was found that both axial and radial actuation close to the combustor inlet were highly efficient. 

As demonstrated by this brief review, the control of the PVC is, on the one hand, subject to research due to its high relevance for different technical applications and, on the other hand, poses significant fundamental questions on the analysis and control of coherent structures in turbulent flows based on global linear stability theory. 

In this work, we address two central questions: Firstly, we investigate where the PVC is most receptive to open-loop forcing within a fully turbulent flow. In contrast to previous work, we focus our attention to the flow region far upstream of the vortex breakdown bubble, which is expected to be most influential for the formation of the PVC, as indicated by previous studies \citep{Kuhn2016,Tammisola2016,Lueckoff2018,Kaiser2018}. Moreover, we validate the receptivity maps using active flow control, which directly leads to the second, more fundamental questions posed in this work: In what extend can the adjoint global mode based on the mean of a highly turbulent flow be used to guide and interpret flow control. Inspired by the work of \cite{Meliga2016a}, we directly compare the results from flow control to the adjoint global modes. In extension to their work, we consider experimental data of a highly three-dimensional flow and the impact of active flow control. Such a setup provokes significant interactions between the applied actuation, the global mode and the mean flow.

To answer both of these open questions, detailed experimental measurements are conducted in a generic turbulent swirling jet setup and an upstream inflow duct of constant cross-section, similar to the setup of \cite{Kuhn2016}. In relation to combustion research, the geometry resembles a generic combustion chamber, including a generic mixing section (the duct). In order to experimentally validate the receptivity determined from the adjoint LSA, zero net mass flux actuation (ZNMF) is employed inside the duct at different axial locations. The required forcing input for synchronization of the actuation and the PVC then quantifies the receptivity of the PVC at the selected actuator positions. The experimental results are compared to the results of the adjoint LSA. In addition, a thorough study of the physical mechanisms leading to synchronization is conducted.

The present paper is structured as follows: first, the experimental methods and setup are introduced. Then, the flow at natural, non-forced conditions is examined in detail, including a global direct and adjoint LSA. The results of the mean flow analyses are discussed. Subsequently, the impact of open-loop forcing and synchronization are evaluated and compared to the results of the adjoint LSA. In the conclusion, key observations and findings are summarized and the results are critically reviewed.

\section{Experimental methods}\label{sec:experimental_methods}

In this section, the experimental setup of the swirling jet facility including the ZNMF actuator is presented. Furthermore, the utilized measurement techniques are explained.

\begin{figure}
    \centering
    \includegraphics[width=0.8\textwidth]{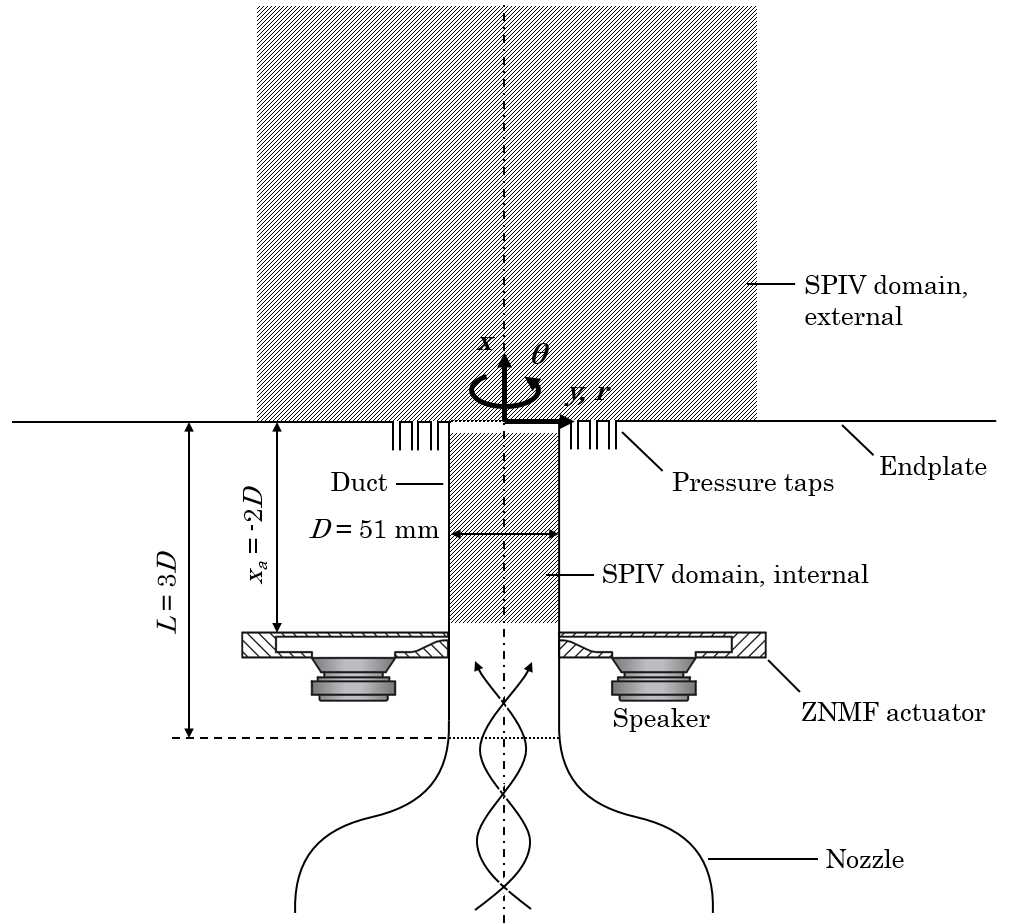}
	\caption{Experimental setup, sectional view, with stereoscopic particle image velocimetry (SPIV) domains and actuator position $x_a/D = -2$ (not to scale)}
    \label{fig:setup_sectionalview}
\end{figure}

A sketch of the overall experimental setup is shown in figure~\ref{fig:setup_sectionalview} in a sectional view. The swirling flow is depicted moving from bottom to top. The swirl is introduced upstream of the nozzle by radial vanes with continuously adjustable vane angles (further details in \cite{Kuhn2016}). Downstream of the swirler, the flow passes through the nozzle with an area contraction ratio of $9$ to $1$ and then enters a duct of constant diameter $D = \SI{51}{\mm}$ and of length $L = \SI{153}{\mm}$, which equals three duct diameters. The flow then exits the duct and emanates into unconfined ambient air. The exit plane is bordered by an endplate.

The used coordinate system originates at the intersection of the exit plane with the duct axis. The $x$-coordinate coincides with the jet axis and is positive in the main flow direction with the $y$- and $z$-coordinate being orthogonal. In this study, the Cartesian coordinates $y$ and $z$ are used interchangeably with the cylindrical coordinates $r$ (radius) and $\theta$ (azimuth). The duct represents the generic mixing section of a combustor for premixed combustion while the area downstream of the duct corresponds to the flow field inside a combustion chamber.

All experiments are conducted for three different mass flow rates at $\dot{m} = 37.5$, $50$ and $\SI{75}{\kg\per\hour}$. These correspond to Reynolds numbers of $\Rey = u_0 D / \nu = 15 \: 000$, $20 \: 000$ and $30 \: 000$ based on the bulk velocity $u_0 = \dot{m} / (\rho \pi D^2 / 4)$ at the duct exit. A constant swirl number was ensured for all three non-forced cases. The swirl number is defined as the ratio of axial flux of azimuthal momentum $G_\theta$ to the axial flux of axial momentum $G_x$ according to \cite{Chigier1964},
\begin{align}
    S = \frac{G_\theta}{D/2 \, G_x}
    \label{eqn:swirlnumber}
\end{align}
with
\begin{align}
    G_\theta \; &= \int_0^\infty \overline{u} \, \overline{w}r^2 \mathrm{d}r\\
    G_x \; &= \int_0^\infty \left( \overline{u}^2 - \overline{w}^2/2 \right) r \mathrm{d}r.
\end{align}
The swirl number $S$ in equation~\ref{eqn:swirlnumber} is calculated close to the duct exit from $x/D = 0.05$ to $0.15$ and the values are then averaged, obtaining a swirl number of $S = 1.18$.  This corresponds to vane angles of $\mu = \SI{64.4}{\degree}$, $\SI{63}{\degree}$ and $\SI{62}{\degree}$ for the respective Reynolds numbers, which was verified by preliminary laser doppler velocimetry measurements.

The critical swirl number for the PVC to occur was determined for $\Rey = 15 \, 000$. In general, it is difficult to precisely determine the bifurcation point in highly turbulent conditions as is the case here. Intermittency of the PVC is treated as an indicator that the flow is at least very close to the bifurcation point, since the oscillations are then likely to be merely noise-driven only \citep{Noiray2013}. Based on this notion, the critical swirl number is estimated as follows: SPIV and pressure measurements are conducted for a range of different vane angles between $\mu = 64.4^\circ$ and $51.4^\circ$ in steps of $1^\circ$. For every case the swirl number is calculated as just described above in equation~\ref{eqn:swirlnumber}. Additionally, a short-time Fourier transform in time is conducted on the recorded pressure signals for the azimuthal wavenumber $m=1$ with a rectangular window with a size of $1$~s. Below the critical swirl number, the PVC occurs only intermittently and the short-time Fourier spectra are characterized by the disappearance and reappearance of a dominant peak. Above the critical swirl number, the PVC is present over the entire length of the measurement ($10$~s) and the peak appears to be steady at one frequency.

\begin{figure}
	\begin{minipage}[t]{0.44\linewidth}
		\includegraphics[width=\textwidth]{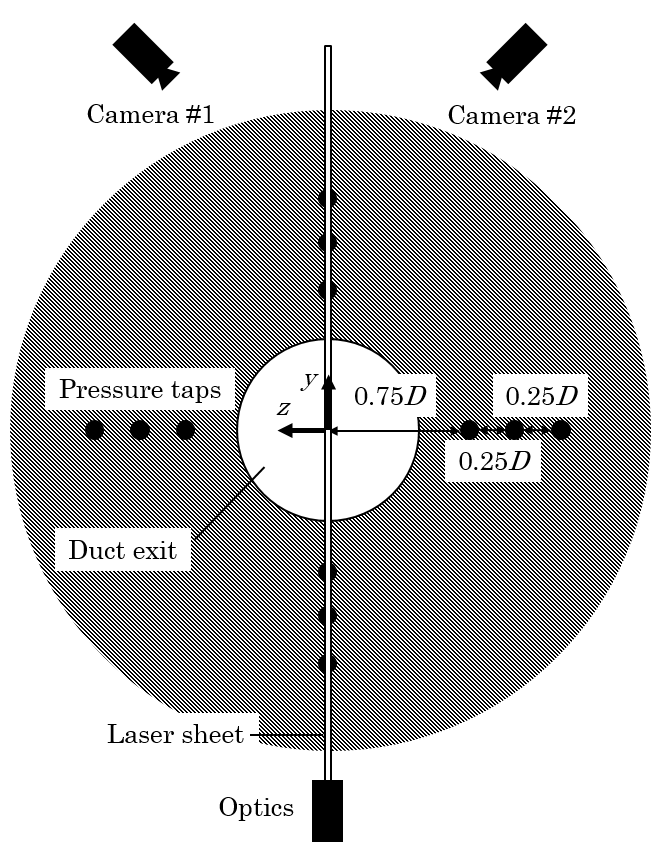}
      	\caption{Experimental setup, top view showing endplate, with pressure tap positions, laser sheet and camera arrangement (not to scale)}\label{fig:setup_topview}
    \end{minipage}
    \hfill
    \begin{minipage}[t]{0.49\linewidth}
		\includegraphics[width=\textwidth]{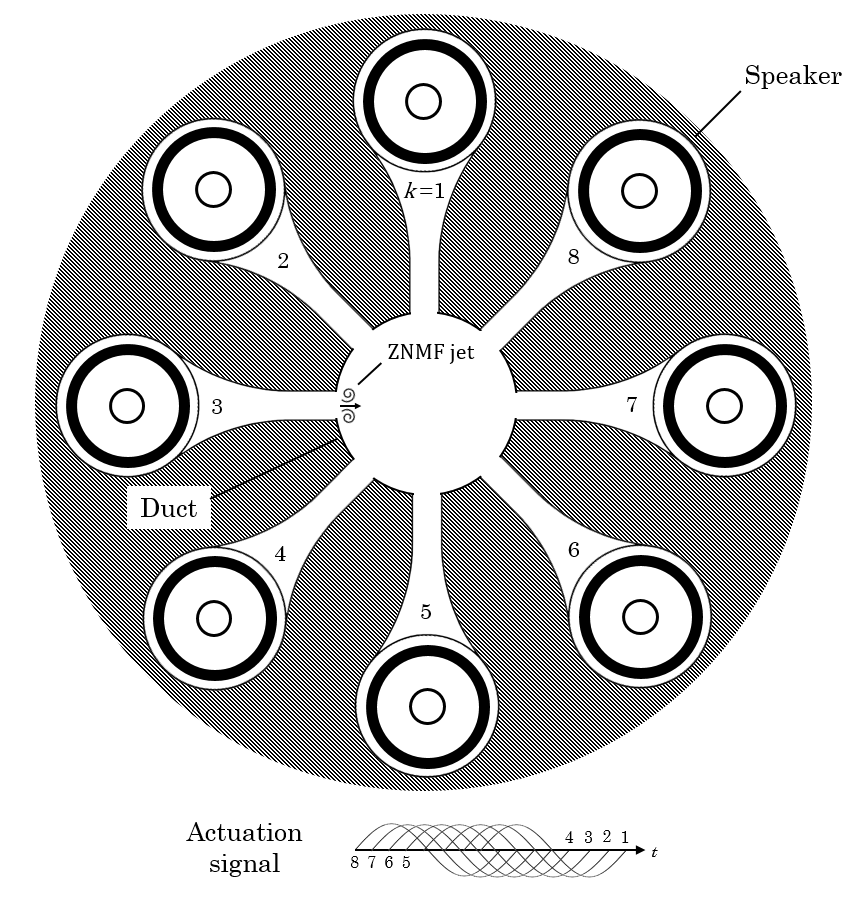}
		\caption{Experimental setup, top view, cross-section of actuator plane (not to scale)}\label{fig:setup_actuator}
    \end{minipage}
\end{figure}

All three components of the velocity field are captured by stereoscopic particle image velocimetry (SPIV), for the $x_a/D = -2$ configuration inside and downstream of the duct, for $x_a/D = -0.5$ only downstream of the duct. This allows for the characterization of the non-forced flow field in general and the forced flow field for these particular actuator positions. A Quantel EverGreen double-pulse Nd:YAG laser with a wavelength of $\SI{532}{\nm}$ and light sheet optics is used to generate a laser sheet of approximately $\SI{2}{\mm}$ thickness. Two PCO 2000 CCD cameras in forward scattering configuration are used for image recording, as illustrated in figure~\ref{fig:setup_topview}. A total number of $N = 1000$ double images per measurement is recorded at a sampling frequency of $\SI{5.89}{\Hz}$. The SPIV measurements are not time-resolved with regard to the PVC. The pulse distances are set according to the respective resulting out-of-plane velocities which depend on the Reynolds number. Flow seeding is achieved with an aerosol of liquid Di-Ethyl-Hexyl-Sebacat. The snapshots with a resolution of $2048 \times 2048$ pixel are post-processed with PIVTEC PIVview. The images are de-warped with the aid of a multi-level calibration target \citep{Willert1997} including disparity correction \citep{Wieneke2005}. Standard digital PIV processing \citep{Willert1991} with an iterative multigrid cross-correlation scheme \citep{Soria1996}, a final window size of $32 \times 32$ pixel, a window overlap of $50\%$ and a subpixel peak fit are employed.

The SPIV domains are separated into an external domain downstream of the duct and an internal domain inside the duct as shown in figure~\ref{fig:setup_sectionalview}. Both domains are not recorded simultaneously. For the internal domain, the light sheet is introduced from downstream direction of the duct exit plane. Due to reflections in the actuator plane, the effectively resolved internal SPIV domain only reaches upstream to $x/D \approx -1.75$. Furthermore, there is an unresolved section between the internal and external domain due to the finite thickness of the endplate ranging from $x/D \approx -0.08$ to $0$. For the mean flow, the separated domains are merged by linearly interpolating the unresolved section. In case of the POD modes, a similar approach is performed by calculating the modes for each domain separately, then matching their phases and finally linearly interpolating between the modes. Due to the rather large extent of both SPIV domains, the resulting camera resolution is insufficient to accurately capture the boundary layers. The missing velocities very close to the walls are therefore extrapolated from the resolved domains onto a no-slip condition $u = v = w = 0$ on the walls.

The natural frequency of the PVC is extracted via time-resolved pressure measurements at a sampling rate of $\SI{50}{\kHz}$, obtained through pressure taps on the endplate. They are arranged concentrically around the duct exit as displayed in figure~\ref{fig:setup_topview}. The differential pressure sensors connected to the taps can resolve pressure fluctuations as small as $\SI{0.1}{\Pa}$. The fluctuating parts of the $j$th sensors $p'_j(t) = p_j(t) - \overline{p_j}$ are decomposed into azimuthal Fourier modes with:
\begin{align}
	\hat{P}_m(t) = \sum_{j=1}^4 p_j'(t) \cdot \expo^{-\imag j m \cdot 2\pi/4}
\end{align}
where $m = 0$, $1$ and $2$ refer to the azimuthal wave numbers that can be resolved with a total amount of four sensors per circumference. The modes are then transformed into a power spectrum with Welch's method \citep{Welch1967} combined with a Hann windowed sliding average at $50\%$ overlap. For obtaining the instantaneous phase angle of the PVC, the time signal of azimuthal mode $m=1$, i.e. $\hat{P}_1(t)$, is filtered first with a Butterworth bandpass filter of order $20$. The filtered signal is then used to calculate the phase angle via:
\begin{align}
	\varphi_\textrm{PVC}(t) = \arctan\left( \frac{\Imag(\hat{P}_1(t)\rvert_{\textrm{filt}})}{\Real(\hat{P}_1(t)\rvert_{\textrm{filt}})} \right) .
\end{align}

\subsection{Open-loop control and synchronization studies}
The PVC is a self-excited oscillator. The oscillatory motion is nearly harmonic with a characteristic natural frequency. When the system is harmonically perturbed or forced by open-loop control, a phenomenon called frequency synchronization may occur where the system
stops to oscillate at its natural frequency and starts to oscillate at the forcing frequency instead. For this, the forcing frequency needs to be sufficiently close to the natural frequency of the system, and the forcing amplitude needs to be of sufficient magnitude. In this work, only 1:1 forced synchronizations are of interest. This means, that the PVC and the forcing are unidirectionally coupled oscillators: the forcing influences the PVC, but not vice versa; and the synchronized frequency of the PVC is equal to the forcing frequency instead of a (reciprocal) integer multiple of the forcing frequency \citep{Balanov2008}.

In order to force the PVC with the goal of reaching synchronization, eight ZNMF actuation units are used, capable of exciting azimuthal modes of $m=0$ to $\pm 4$. Each unit consists of one loudspeaker with a rated input power of $\SI{15}{\W}$ connected by small channels to the duct, as sketched in figure~\ref{fig:setup_actuator}. Each stroke period of the speaker creates a ZNMF jet at the slot exit of the actuator channel with a rectangular cross-section of $\SI{1}{\mm} \times \SI{22}{\mm}$. Addressing each speaker with a different phase lag at a specific time allows periodic excitation. The actuation is applied at the azimuthal wavenumber of $m=1$, equal to the azimuthal wavenumber of the PVC. The input actuation signal $A_k$ of speaker no. $k$ is, therefore, given by
\begin{align}
    A_k = A_f \sin{\left( 2\pi \left[ f_f t + \frac{k-1}{8} \right] \right)}
\end{align}
with $A_f$ being the forcing amplitude and $f_f$ being the forcing frequency.

All speakers are calibrated at no-flow conditions with a microphone placed at $r = 0$ in the actuation plane for the relevant frequency range in order to produce repeatable sound pressure levels. Moreover, preliminary hot-wire measurements at no-flow conditions show that the transfer function of input voltage to output peak velocity is linear within the utilized frequency and amplitude ranges. Nonetheless, amplitudes in the actuation experiments are still given in voltage since the peak velocities at no-flow conditions are likely to be overestimated compared to flow conditions, thus not being directly proportional. Therefore, forcing and synchronization amplitudes are simply quantified by the input voltage of the speakers.

Five actuation positions are tested for the open-loop control studies. Measured from the exit plane of the duct these are: $x_a/D = -2$, $-1.5$, $-1$, $-0.75$ and $-0.5$. Actuation is applied within the range of $\pm 10\%$ of the natural frequency. The maximum frequency increments are $\pm 2.5\% f_n$ or lower between the selected forcing frequencies $f_f$. One of these five actuation positions is exemplarily illustrated in figure~\ref{fig:setup_sectionalview}. The experimental parameters are summarized in Tab.~\ref{tab:exp_parameters}.

\renewcommand{\arraystretch}{1.5}
\addtolength{\tabcolsep}{0.3cm}
\begin{table}
	\centering
  	\begin{tabular}{c c c c c | c}
  	$\dot{m}$ [$\si{\kg\per\hour}$] & $u_0$ [$\si{\m\per\s}$] & $\Rey$ [$-$] & $\mu$ [$\si{\degree}$] & $S$ [$-$] & $x_a/D$ [$-$] \\
  	$37.5$ & $4.3$ & $15 \: 000$ & $64.4$ & $1.18$ & \multirow{3}{*}{\shortstack[l]{$-2$, $-1.5$,\\$-1$, $-0.75$,\\$-0.5$}} \\
  	$50$ & $5.7$ & $20 \: 000$ & $63$ & $1.18$ & \\
  	$75$ & $8.6$ & $30 \: 000$ & $62$ & $1.18$ & \\
  	\end{tabular}
    \caption{Overview of experimental parameters.}
  	\label{tab:exp_parameters}
\end{table}
\addtolength{\tabcolsep}{-0.3cm}

A viable criterion whether synchronization has been reached can be defined by considering the phase angle difference $\Delta\varphi$ between the instantaneous phase of the PVC $\varphi_\textrm{PVC}$ and the phase of the forcing $\varphi_f$:
\begin{align}
	\Delta\varphi = \varphi_\textrm{PVC} - \varphi_f + \Delta\varphi_0
\end{align}
with $\Delta\varphi_0$ as the initial phase difference at $t=0$, arbitrarily set to $\Delta\varphi_0 = 0$ here. Synchronization is established when:
\begin{align}
	\Delta\varphi \approx 0\textrm{,} \; \textrm{for} \; t \to \infty
    \label{eqn:lockin}
\end{align}
i.e. when the PVC and forcing frequency are approximately equal for the time of measurement of \SI{10}{\s}. Exact equality is not possible in this highly turbulent setup. The forcing amplitude where synchronization is reached is called synchronization amplitude.

\section{Theoretical methods}
The theoretical methods employed in this work are explained in this section. On the one hand, proper orthogonal decomposition is used to extract the PVC as an empirical mode directly from the time series of the velocity fields. On the other hand, global direct LSA is used to calculate the eigenmode of the PVC from the mean flow. Furthermore, global adjoint LSA is employed in order to characterize the receptivity and structural sensitivity of the PVC.

One common method to describe coherent structures is the triple decomposition technique \citep{Reynolds1972}:
\begin{align}
	\vec{q}(\vec{x},t) = \overline{\vec{q}}(\vec{x}) + \widetilde{\vec{q}}(\vec{x},t) + \vec{q}''(\vec{x},t),
	\label{eqn:tripledecomposition}
\end{align}
where $\vec{q}$ is an arbitrary space and time dependent quantity of the flow, $\overline{\vec{q}}$ is the time-average or mean part, $\widetilde{\vec{q}}$ is the coherent fluctuation part (time-average subtracted from phase-average) and $\vec{q}''$ is the stochastic fluctuation part. By that, the triple decomposition is basically a refined version of the Reynolds decomposition where all turbulent fluctuations are absorbed into a single fluctuation part $\vec{q}' = \widetilde{\vec{q}} + \vec{q}''$. The coherent part quantifies all those structures in the turbulence spectrum which are periodic in time and space.

\subsection{Empirical modes by proper orthogonal decomposition}
The PVC is a coherent structure and, therefore, typically described with the triple decomposition technique. For extracting dominant coherent velocity fluctuations from a turbulent velocity field, the POD is a well tested technique \citep{Berkooz1993}. In the POD, an ensemble of $N$ velocity fields is projected onto an orthogonal $N$-dimensional basis that maximizes the turbulent kinetic energy for any subspace. Thus, the POD provides an optimal set of spatial modes with which the energy containing structures can be described in the most efficient way with as few modes as possible. The POD of a velocity field $\vec{u}$ is determined with:
\begin{align}
	\vec{u}(\vec{x},t) = \overline{\vec{u}}(\vec{x}) + \sum_{j=1}^N a_j(t) \vec{\Phi}_j(\vec{x}) + \vec{u}_{\textrm{res}}(\vec{x},t)
\end{align}
by minimizing the residual $\vec{u}_{\textrm{res}}$. The $a_j(t)$ are the POD time coefficients, while the $\vec{\Phi}_j(\vec{x})$ are the spatial POD modes. In case of SPIV, $N$ is the number of snapshots. The POD modes $\vec{\Phi}_j$ provide the spatial shape of the mode and the time coefficients $a_j$ provide the time dependent amplitude of the modes. Furthermore, the mean specific turbulent kinetic energy of the associated modes can be calculated via $\lambda_j = \overline{a_j^2}$. The POD modes are energy ranked such that $\lambda_1 \geq \lambda_2 \geq \ldots \geq \lambda_N$. In the case of the PVC, the global mode is typically captured by two modes with the highest energy content at similar energy level. Examining the phase portraits of the corresponding time coefficients reveals whether the modes are indeed periodic. When two of such modes occur, the coherent velocity fluctuation of the PVC can be reconstructed by:
\begin{align}
	\widetilde{\vec{u}}(\vec{x},t) = \Real \left\{ \overline{\sqrt{a_1^2 + a_2^2}} \, \left( \vec{\Phi}_1(\vec{x}) + \imag \vec{\Phi}_2(\vec{x}) \right) e^{-2\pi \imag f_\textrm{PVC} t} \right\}
    \label{eqn:pod_superposition}
\end{align}
where $\Real$ is the real part and $f_\textrm{PVC}$ is the frequency of the global mode. Since the SPIV measurements are not time-resolved with regard to the PVC oscillations, the frequency cannot be obtained via a spectral analysis of the time coefficients. Instead, the frequency is determined from the time-resolved pressure measurements.

\subsection{Eigenmodes by global direct linear stability analysis} \label{sec:global_direct_lsa}
The LSA in turbulent flows is used to obtain modes of coherent velocity fluctuation. In incompressible flows, the governing equations for the hydrodynamic LSA are derived from the incompressible Navier--Stokes equations and the incompressible continuity equation. The triple decomposition ansatz (equation~\ref{eqn:tripledecomposition}) is substituted into both equations and both are time-averaged and phase-averaged. By subtracting the time-averaged set of equations from the phase-averaged set of equations, the governing equations for the coherent velocity fluctuations are obtained \citep{Reynolds1972}:
\begin{align}
	\frac{\partial \widetilde{\vec{u}}}{\partial t} + (\widetilde{\vec{u}} \cdot \nabla) \, \overline{\vec{u}} + (\overline{\vec{u}} \cdot \nabla) \, \widetilde{\vec{u}} &= - \frac{1}{\rho} {\nabla \widetilde{p}} + \nabla \cdot (\, \nu \, (\nabla + \nabla^{\top}) \, \widetilde{\vec{u}} \,) \notag \\ &\phantom{{}=} - \nabla \cdot (\; \tens{\tau}_R \: + \underbrace{\tens{\tau}_N}_{\approx 0} ) \label{eqn:coherent_nse} \\
    \nabla \cdot \widetilde{\vec{u}} &= 0  .
\end{align}

The term $\tens{\tau}_N$ describes the nonlinear interactions of the perturbation with its higher harmonics. Under the assumption that these interactions are very weak this term is neglected in the following. The term $\tens{\tau}_R = \langle \vec{u}'' \vec{u}'' \rangle - \overline{\vec{u}'' \vec{u}''} = \widetilde{\vec{u}'' \vec{u}''}$ describes the fluctuation of the stochastic Reynolds stresses due to the passage of a coherent perturbation \citep{Reynolds1972}. This term has to be modeled in order to close equation~\ref{eqn:coherent_nse}. In the context of the LSA in swirling flows with a PVC, it is well established to use Boussinesq's eddy viscosity model as a closure \citep{Reau2002b,Reau2002a,Crouch2007,Paredes2016,Tammisola2016,Rukes2016,Kaiser2018}. This approach is used here as well. It is assumed that the coherent fluctuations of the turbulent kinetic energy are small enough to be negligible \citep{Reynolds1972}. With that assumption the Reynolds stresses are expressed as
\begin{align}
	\tens{\tau}_R = \widetilde{\vec{u}'' \vec{u}''} = - \nu_t ( \nabla + \nabla^{\top} ) \widetilde{\vec{u}} .
    \label{eqn:nut_def}
\end{align}
The unknown eddy viscosity is calculated from the known velocity field of the experiment. As the turbulence of the swirling jet is highly anisotropic \citep{Rukes2016}, the approach in equation~\ref{eqn:nut_def} yields six independent eddy viscosities. A reasonable compromise among the six eddy viscosities can be achieved by using a least-square fit over all resolved stochastic Reynolds stresses \citep{Ivanova2013}:
\begin{align}
	\nu_t = \frac{\langle -\overline{\vec{u}'' \vec{u}''} + 2/3 \: k''  \vec{I} ,\: \overline{\vec{S}} \rangle_\textrm{F}}{2 \langle \overline{\vec{S}} ,\: \overline{\vec{S}} \rangle_\textrm{F}}
	\label{eqn:nut_lsq}
\end{align}
where $\langle \: \cdot \:,\: \cdot \: \rangle_\textrm{F}$ is the Frobenius inner product, $k''$ is the turbulent-stochastic kinetic energy, $\vec{I}$ is the identity tensor and $\overline{\vec{S}} = 1/2 (\nabla + \nabla^{\top}) \overline{\vec{u}}$ is the mean strain rate tensor. The eddy viscosity is then simply added to the kinematic viscosity in equation~\ref{eqn:coherent_nse} to form an effective viscosity $\nu_\mathrm{eff} = \nu + \nu_t$. 

\begin{figure}
	\centering
  	\includegraphics[width=0.4\textwidth]{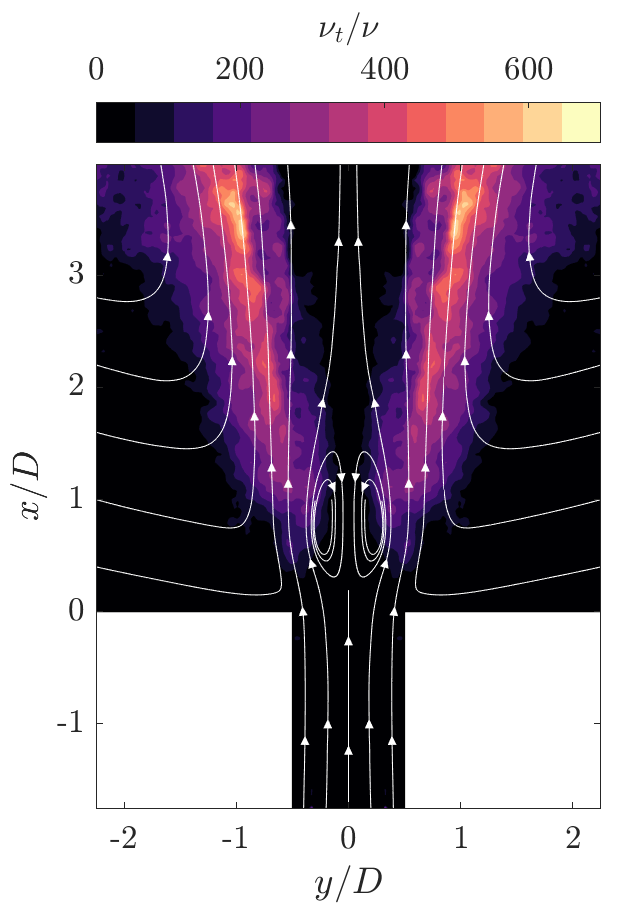}
	\caption{Eddy viscosity $\nu_t$ normalized with molecular viscosity $\nu$, calculated via least-square fit (equation~\ref{eqn:nut_lsq}), $\Rey = 20 \: 000$}
    \label{fig:nut_lsq}
\end{figure}

Figure~\ref{fig:nut_lsq} shows the eddy viscosity obtained with equation~\ref{eqn:nut_lsq}. The eddy viscosity is particularly high in the inner and outer shear layers of the swirling jet with increasing eddy viscosity in downstream direction. This is adequate to describe the downstream dissipation of the coherent structures that occur due to the interaction with small-scale turbulence. Due to measurement noise, calculation of the discretized gradients in equation~\ref{eqn:nut_lsq} is exposed to that same noise. In order to ameliorate the noisy gradient fields, a box blur low-pass filter is used with a discrete kernel size of $3$ corresponding to every neighboring point, prior to the calculation of the eddy viscosity. Very small negative values only occur at $4 \cdot 10^{-3} \: \%$ of the nodes of the entire domain and are confined to the boundary layer flow inside the duct. All of these negative values are rigorously set to $0$.

The linearized Navier--Stokes and continuity equation for the coherent fluctuation become:
\begin{align}
	\frac{\partial \widetilde{\vec{u}}}{\partial t} + (\widetilde{\vec{u}} \cdot \nabla) \, \overline{\vec{u}} + (\overline{\vec{u}} \cdot \nabla) \, \widetilde{\vec{u}} &= - \frac{1}{\rho} {\nabla \widetilde{p}} + \nabla \cdot (\, \nu_\mathrm{eff} \, (\nabla + \nabla^{\top}) \, \widetilde{\vec{u}} ) \label{eqn:coherent_lnse} \\
	\nabla \cdot \widetilde{\vec{u}} &= 0 .\label{eqn:continuity}
\end{align}
These equations can be rewritten into
\begin{align}
	\vec{\mathcal{B}} \frac{\partial \widetilde{\vec{q}}}{\partial t} - \vec{\mathcal{A}} \widetilde{\vec{q}} = 0
	\label{eqn:coherent_lnse_operatorform}
\end{align}
where $\vec{\mathcal{A}}$ and $\vec{\mathcal{B}}$ represent the operators of equation~\ref{eqn:coherent_lnse} and \ref{eqn:continuity} and $\widetilde{\vec{q}} = [\widetilde{\vec{u}}, \widetilde{p}]^{\top}$ is the combined vector of the velocities and pressure.

The global LSA examines flows which are inhomogeneous in two or three spatial dimensions. These are typically named biglobal and triglobal LSA, respectively \citep{Theofilis2003}. Within the scope of the PVC, a biglobal analysis suffices due to the homogeneity in azimuthal direction. Equation~\ref{eqn:coherent_lnse_operatorform} is solved with a normal mode ansatz in cylindrical coordinates:
\begin{align}
	\widetilde{\vec{q}}(\vec{x},t) = \Real \left\{ \hat{\vec{q}}(x,r) e^{\imag (m\theta - \omega t)} \right\}
    \label{eqn:normalmode_ansatz}
\end{align}
where $m$ is the azimuthal wave number and $\omega$ is the complex angular frequency. Since the PVC is a single-helical mode, the azimuthal wave number is set to $m = 1$. Discretization and rearrangement lead to a generalized eigenvalue problem with $\omega$ as the eigenvalue, which can be written in the form of
\begin{align}
	\vec{A} \hat{\vec{q}} = \omega \vec{B} \hat{\vec{q}}
    \label{eqn:egv_problem}
\end{align}
in which $\vec{A}$ and $\vec{B}$ are the discretized operators of equation~\ref{eqn:coherent_lnse_operatorform}, including equation~\ref{eqn:normalmode_ansatz}. Solving equation~\ref{eqn:egv_problem} provides the eigenmodes $\hat{\vec{q}}$, each accompanied with one complex eigenvalue $\omega$, respectively.

It consists of a real part $\Real(\omega)$ that corresponds to the angular frequency of the mode and an imaginary part $\Imag(\omega)$ that corresponds to the growth rate of the mode. The mode is stable when $\Imag(\omega) < 0$, marginally stable when $\Imag(\omega) = 0$ and unstable when $\Imag(\omega) > 0$. In the eigenspectrum, an oscillator mode at limit cycle, such as the PVC, is expected to be represented by an eigenvalue isolated from any continuous eigenvalue branch and ideally marginally stable ($\Imag(\omega) = 0$) since the instability neither grows nor decays \citep{Barkley2006}. With that criterion and the known frequency from the experiment, the PVC mode can be identified and written as:
\begin{align}
	\widetilde{\vec{u}}(\vec{x},t) = \Real \left\{ \hat{\vec{u}}(x,r) e^{\imag (\theta - \Real(\omega) t)} \right\} .
\end{align}

\begin{figure}
    \centering
    \includegraphics[width=0.6\textwidth]{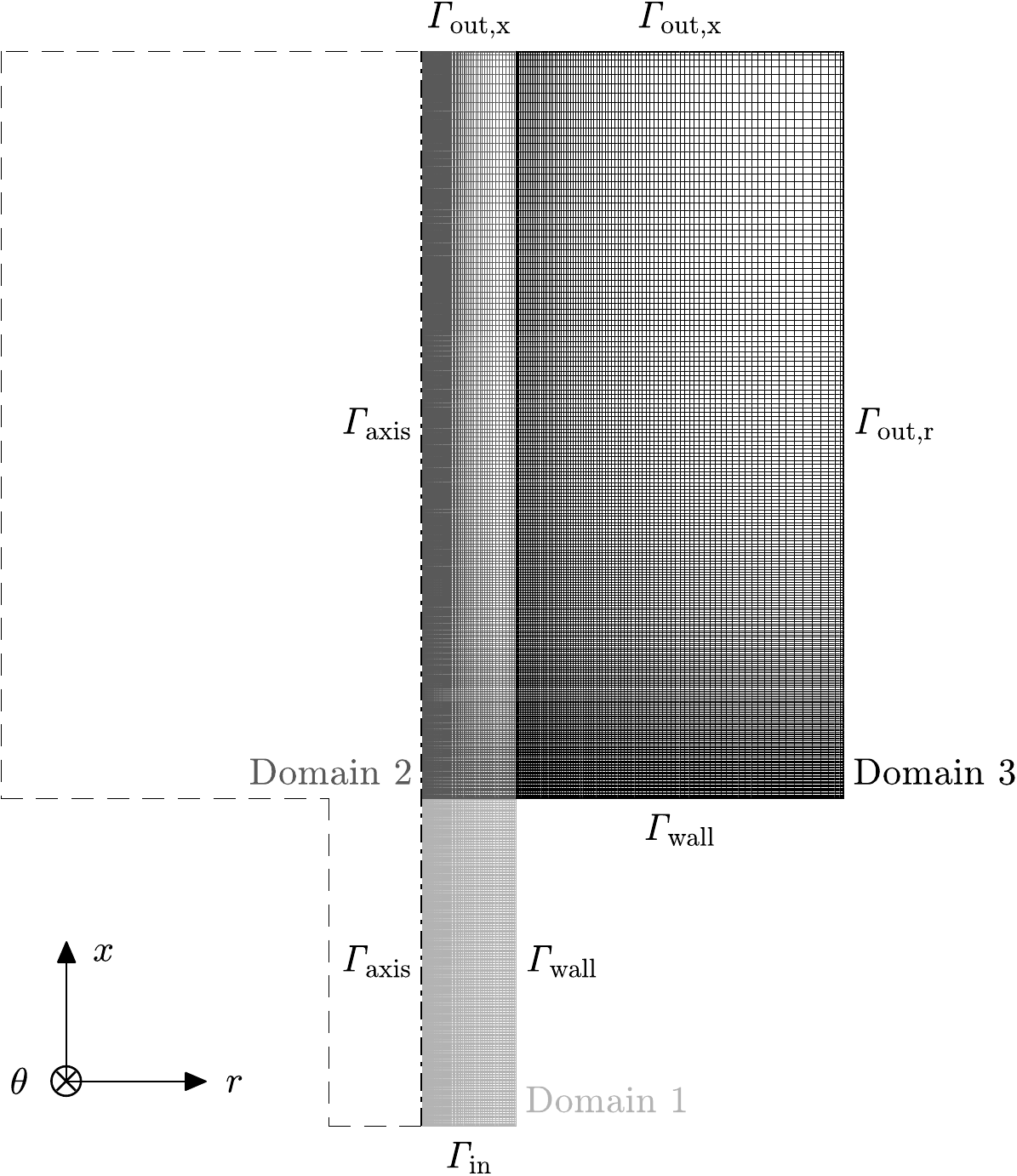}
	\caption{Mesh grid of the discretized domain for the global LSA including the boundaries $\Gamma$, the three rectangular subdomains with conforming interfaces are marked in terms of color}
    \label{fig:mesh}
\end{figure}

For discretizing and solving the generalized eigenvalue problem (equation~\ref{eqn:egv_problem}) a Fortran code by \cite{Paredes2014} is employed that uses the Arnoldi algorithm. The domain is decomposed into three rectangular subdomains with conforming interfaces \citep{DemaretDeville1991}. A high-order finite difference scheme with nonuniformly distributed nodes is used for the discretization of the axial and radial directions which is able to provide a converged solution at comparatively low mesh resolution \citep{Paredes2013a}. Figure~\ref{fig:mesh} shows the grid for the final resolution with a total number of $70104$ grid nodes. The three subdomains are marked in terms of color. The grid spacing in axial and radial direction is particularly small inside the duct (domain~$1$) and in the vicinity of the duct exit (lower part of domain~$2$ and $3$) to accurately capture the PVC mode. The grid spacing is then expanded in downstream and far field direction.

Homogeneous Neumann boundary conditions for velocity and pressure are imposed at the inlet $\Gamma_\textrm{in}$. This is necessary since the coherent perturbations are low but not zero at the upstream boundary as observed in the experiments (figure~\ref{fig:pod_modes_re20k}). For the same reason, homogeneous Neumann boundary conditions for the velocity and pressure are set at the axial outlet $\Gamma_\textrm{out,x}$ in the direct LSA case. Here, the computational domain is truncated at a position with remaining nonzero perturbations. In the adjoint LSA case, homogeneous Dirichlet conditions are set at the axial outlet $\Gamma_\textrm{out,x}$ since no adjoint perturbations are allowed to enter the domain. At the radial outlet $\Gamma_\textrm{out,r}$, homogeneous Dirichlet boundary conditions are set. For all walls $\Gamma_\textrm{wall}$, homogeneous Dirichlet boundary conditions are imposed for the velocity due to the no-slip and no-penetration condition. For the pressure, no physical boundary conditions exist. However, a compatibility condition from the governing momentum equations can be derived \citep{Theofilis2004}. The homogeneous Dirichlet conditions for the velocity are substituted into the linearized Navier--Stokes equation for the coherent perturbation (equation~\ref{eqn:coherent_lnse}). Knowing that the eddy viscosity is zero per definition on the wall and taking the inner product with the unit normal vector $\vec{n}$ at the respective boundaries, rearrangement leads to
\begin{align}
	\frac{\partial p}{\partial n} = \rho \nu \frac{\partial^2 u_n}{\partial n^2}, \;\; \vec{x} \in \Gamma_\textrm{wall} .
\end{align}
Assuming $\partial^2 u_n / \partial n^2 \approx 0$ provides homogeneous Neumann conditions for the pressure on the walls. Since the global LSA is conducted in a cylindrical coordinate system, the del operator $\nabla$ in equation~\ref{eqn:coherent_lnse} and \ref{eqn:continuity} exhibits a singularity in $r = 0$. To ensure smoothness and boundedness on the centerline, the boundary conditions are set in agreement with \cite{Khorrami1989}. Homogeneous Neumann boundary conditions are set for $\hat{v}$ and $\hat{w}$ on the axis $\Gamma_\textrm{axis}$ whereas homogeneous Dirichlet conditions are set for $\hat{u}$ and $\hat{p}$. All boundary conditions are summarized in Tab.~\ref{tab:mafia_bcs}.

\renewcommand{\arraystretch}{1.35}
\addtolength{\tabcolsep}{0.3cm}
\begin{table}
	\centering
  	\begin{tabular}{c c c c c c}
  	$\Gamma_\textrm{in}$ & $\Gamma_\textrm{out,x}$ & $\Gamma_\textrm{out,x}$ & $\Gamma_\textrm{out,r}$ & $\Gamma_\textrm{wall}$ & $\Gamma_\textrm{axis}$ \\
  	\scriptsize{(direct \&} & \scriptsize{(direct)} & \scriptsize{(adjoint)} & \scriptsize{(direct \&} & \scriptsize{(direct \&} & \scriptsize{(direct \&}\\
  	 \scriptsize{adjoint)} & & &  \scriptsize{adjoint)} &  \scriptsize{adjoint)} &  \scriptsize{adjoint)}\\[0.3cm]
    $\partial \hat{u} / \partial n = 0$ & $\partial \hat{u} / \partial n = 0$ & $\hat{u} = 0$ & $\hat{u} = 0$ & $\hat{u} = 0$ & $\hat{u} = 0$ \\
    $\partial \hat{v} / \partial n = 0$ & $\partial \hat{v} / \partial n = 0$ & $\hat{v} = 0$ & $\hat{v} = 0$ & $\hat{v} = 0$ & $\partial \hat{v} / \partial r = 0$ \\
    $\partial \hat{w} / \partial n = 0$ & $\partial \hat{w} / \partial n = 0$ & $\hat{w} = 0$ & $\hat{w} = 0$ & $\hat{w} = 0$ & $\partial \hat{w} / \partial r = 0$ \\
    $\partial \hat{p} / \partial n = 0$ & $\partial \hat{p} / \partial n = 0$ & $\hat{p} = 0$ & $\hat{p} = 0$ & $\partial \hat{p} / \partial n = 0$ & $\hat{p} = 0$ \\
  	\end{tabular}
    \caption{Boundary conditions of the global LSA}
  	\label{tab:mafia_bcs}
\end{table}
\addtolength{\tabcolsep}{-0.4cm}

\subsection{Receptivity and structural sensitivity by global adjoint linear stability analysis}\label{sec:receptivity_and_SS}
The global adjoint LSA is employed to estimate the receptivity of the PVC. This provides a theoretical prediction where the PVC is most efficiently controlled by periodic forcing. Furthermore, the adjoint LSA enables to calculate the structural sensitivity which reveals where the wavemaker of the PVC is located which is responsible for the continuous self-excitation of the instability.

The discretized operator $\vec{A}$ in equation~\ref{eqn:egv_problem} is typically non-normal, especially in shear flows. This non-normality causes the eigenvectors of the system to be strongly nonorthogonal \citep{Sipp2010}. When an arbitrary vector is expanded in the basis of eigenvectors, a dual basis is required due to this nonorthogonality \citep{Salwen1972,Tumin1984,Tumin1996,Oden2017}. In case of the global LSA, the dual of the direct eigenbasis is the basis of adjoint eigenvectors. These are obtained from the generalized adjoint eigenvalue problem \citep{Luchini2014}
\begin{align}
	\vec{A}^\mathrm{H} \hat{\vec{q}}^+ &= \omega^+ \vec{B}^\mathrm{H} \hat{\vec{q}}^+
    \label{eqn:adj_problem}
\end{align}
where $( \cdot )^\mathrm{H}$ denotes the Hermitian transpose and $( \cdot )^+$ denotes an adjoint quantity (note that $\omega^+ = \omega^\mathrm{H}$). In case of a converged adjoint solution, the adjoint eigenvalue $\omega^+_j$ of the adjoint eigenmode $\hat{\vec{q}}^+_j$ is the complex conjugate of the direct eigenvalue $\omega_j$. By that, the pair of mutual direct and adjoint eigenmodes, $\hat{\vec{q}}_j$ and $\hat{\vec{q}}^+_j$, are identified.

The bases of direct and adjoint eigenvectors form a biorthogonal system \citep{Oden2017}. Let the Hermitian inner product on a complex vector space be defined by $\langle \vec{M} \vec{x}, \vec{y} \rangle = (\vec{M} \vec{x})^\mathrm{H}\vec{y} = \langle \vec{x}, \vec{M}^\mathrm{H} \vec{y} \rangle$. Considering the generalized direct eigenvalue problem of equation~\ref{eqn:egv_problem} and the generalized adjoint eigenvalue problem of equation~\ref{eqn:adj_problem}, the biorthogonality condition can be easily derived and reads \citep{Luchini2014}
\begin{align}
	(\omega_k - (\omega_j^+)^\mathrm{H}) \langle \hat{\vec{q}}^+_j , \vec{B} \hat{\vec{q}}_k \rangle = 0 .
	\label{eqn:biorthogonality}
\end{align}
If $k=j$, the expression inside the round brackets vanishes since $\omega_{k} = (\omega_j^+)^\mathrm{H}$. If $k \neq j$, it directly follows that $\langle \hat{\vec{q}}^+_j , \vec{B} \hat{\vec{q}}_k \rangle = 0$. In other words, all direct and adjoint eigenvectors are pairwise orthogonal with regard to $\vec{B}$, except if $k=j$. This property of biorthogonality can be used (a) to expand the solution of the \textit{inhomogeneous} linearized Navier--Stokes equation for the coherent fluctuation (equation~\ref{eqn:coherent_lnse_operatorform} including a source term on the right-hand side) in order to characterize the receptivity of an instability; and (b) to expand the solution of the \textit{perturbed} generalized eigenvalue problem (equation~\ref{eqn:egv_problem} including small perturbations of the discretized operator $\vec{A}$) in order to localize the wavemaker of an instability.

For (a) characterizing the receptivity to periodic open-loop forcing, a momentum source on the right-hand side of equation~\ref{eqn:coherent_lnse_operatorform} is introduced. The source term is assumed to be helical in azimuthal direction, to be harmonic in time and to not interact with the homogeneous system. Hence, the eigenspectrum is not modified by the source and the source can be interpreted as an open-loop forcing. The initial value problem \ref{eqn:coherent_lnse_operatorform} is then modified and written as
\begin{align}
	\vec{\mathcal{B}} \frac{\partial \widetilde{\vec{q}}}{\partial t} - \vec{\mathcal{A}} \widetilde{\vec{q}} = \vec{\hat{f}} e^{\imag (\theta - \omega_f t)}
	\label{eqn:coherent_lnse_operatorform_inhom}
\end{align}
with $\vec{\hat{f}}$ being the spatial structure of the forcing and $\omega_f$ being the angular forcing frequency. The solution is derived making use of the biorthogonality condition in equation~\ref{eqn:biorthogonality}. The derivation is not shown here for brevity and the interested reader is referred to \cite{Chandler2011} for a complete derivation. Let $\hat{\vec{q}}_1$ be the eigenmode of the PVC and $\hat{\vec{q}}_0$ be the initial condition. It is assumed that the system already oscillates with the PVC eigenmode in natural state, prior to applying open-loop forcing, i.e. $\hat{\vec{q}}_0 = \hat{\vec{q}}_1$. When oscillating close to limit cycle, the decay rate is much smaller than the decay rate of all the remaining modes, $\lvert \Imag (\omega_1) \rvert \ll \lvert \Imag (\omega_{j \neq 1}) \rvert$. Also, the forcing is assumed to be of constant amplitude, thus $\Imag (\omega_f) = 0$. For $t$ becoming very large, the discrete solution of equation~\ref{eqn:coherent_lnse_operatorform_inhom} can then be approximated by
\begin{align}
	\widetilde{\vec{q}} \approx
	\Real \bigg\{ \bigg( 1
	&- \overbrace{ \frac{\langle \hat{\vec{q}}^+_1, \vec{\hat{f}} \rangle}{(\omega_f - \omega_1) \langle \hat{\vec{q}}^+_1, \vec{B} \hat{\vec{q}}_1 \rangle} }^{R_f} \bigg) \hat{\vec{q}}_1 e^{\imag (\theta - \omega_1 t)} \nonumber \\
	&+ \underbrace{ \frac{\langle \hat{\vec{q}}^+_1, \vec{\hat{f}}  \rangle}{(\omega_f - \omega_1) \langle \hat{\vec{q}}^+_1, \vec{B} \hat{\vec{q}}_1 \rangle} }_{R_f} \;\;\: \hat{\vec{q}}_1 e^{\imag (\theta - \omega_f t)} \bigg\} .
	\label{eqn:coherent_lnse_inhom_solution}
\end{align}
The resulting spatial mode shape of the system is not altered by the forcing $\vec{\hat{f}}$ and is the PVC mode $\hat{\vec{q}}_1$ itself. The temporal response is both at the PVC frequency $\omega_1$ and at the forcing frequency $\omega_f$ and proportional to $\lvert R_f \rvert$. $\lvert R_f \rvert$ increases when the spatial structure of the forcing $\vec{\hat{f}}$ approaches the spatial structure of the adjoint PVC mode $\hat{\vec{q}}^+_1$. Additionally, $\lvert R_f \rvert$ increases when the forcing frequency $\omega_f$ approaches the PVC frequency $\omega_1$. According to the Cauchy--Schwarz inequality, $\lvert R_f \rvert \propto \lvert \langle \hat{\vec{q}}^+_1, \vec{\hat{f}} \rangle \rvert \leqslant \lvert \hat{\vec{q}}^+_1 \rvert \lvert \vec{\hat{f}} \rvert$ holds. Therefore, the magnitude of the adjoint mode can be physically interpreted as the receptivity of the system to a given open-loop periodic forcing. Note that for $\vec{\hat{f}} = \vec{0}$, the original PVC mode as an eigensolution of equation~\ref{eqn:coherent_lnse_operatorform}, i.e. $\widetilde{\vec{q}} = \hat{\vec{q}}_1$, is retrieved.

It has to be noted that the receptivity to open-loop forcing defined by equation~\ref{eqn:coherent_lnse_inhom_solution} is only valid in the sense it was mathematically introduced. The most severe restriction is the assumption that open-loop actuation only acts as a source term in the coherent momentum equation, without interacting with the PVC mode itself and without modifying the mean flow. As will be shown later, this assumption is not valid in general and nonlinear interaction and mean flow modifications do indeed occur in the process of synchronization. The adjoint theory has, therefore, its limits within this assumption, and will only give a gradient-like estimation but not a complete description of the synchronization process.

In order to locate the wavemaker of the PVC instability, the region of strongest intrinsic feedback is sought. The feedback can be modeled as a source term in equation~\ref{eqn:egv_problem}, similar to equation~\ref{eqn:coherent_lnse_operatorform_inhom} but this time proportional to the perturbation $\widetilde{\vec{q}}$ itself such that
\begin{align}
    \vec{\mathcal{B}} \frac{\partial \widetilde{\vec{q}}}{\partial t} - \vec{\mathcal{A}} \widetilde{\vec{q}} = \vec{\mathcal{C}} \widetilde{\vec{q}}.
    \label{eqn:structural_perturbation}
\end{align}
This results in a perturbed generalized eigenvalue problem due to a structural perturbation of the linear operator $\vec{\mathcal{A}}$. This is equivalent to a modification of the base or mean flow and results in a change of the eigenvalue. Considering only spatially localized structural perturbations, the region where the resulting eigenvalue change is the highest is the region where the intrinsic feedback is the strongest. This eigenvalue change can be calculated at first order and a bound of this change at every position is obtained by utilizing the Cauchy--Schwarz inequality. The upper bound is typically called structural sensitivity and, in $L^2$ norm, reads \citep{Giannetti2007}
\begin{align}
    \Lambda(x, r) = \lvert \hat{\vec{u}}^+(x, r) \rvert \lvert \hat{\vec{u}}(x, r) \rvert .
    \label{eqn:structural_sensitivity}
\end{align}
The position of the wavemaker, which drives the self-excitation of the instability, corresponds to the region where the structural sensitivity is significantly high.

\section{The precessing vortex core in natural state}
This section examines the baseline case of the PVC without applied open-loop control. At first, the mean flow is shortly characterized. Subsequently, the PVC is investigated by inspecting the results from Fourier and proper orthogonal decomposition as well as global LSA. The main features and mechanisms of the PVC are discussed and the agreement between POD and LSA is evaluated. Furthermore, the receptivity and structural sensitivity of the PVC is examined via the global adjoint LSA in order to reveal where the PVC can be efficiently manipulated and where the wavemaker is located.

By choosing the normalization parameters bulk velocity $u_0$ and duct diameter $D$, the mean flow field can be nondimensionalized. In case of the coherent structures, the normalization is performed with the maximum magnitude of the coherent velocity vector. Using this normalization, the results for all three Reynolds numbers $\Rey = 15 \: 000$, $20 \: 000$ and $30 \: 000$ exhibit a strong similarity. Furthermore, the Strouhal number is also nearly the same for all three cases. Therefore, for brevity, most of the following discussions can be reduced to considering the results at one Reynolds number only. Unless otherwise stated, $Re=20 \: 000$ is selected.

Due to the Reynolds number independence of the flow it is to be expected that the following results apply to flows at even higher Reynolds numbers as well. Therefore, the PVC can be treated as an inviscid instability. Only compressibility effects may play an increasing role when the Mach number is equally increased as the Reynolds number.

\subsection{Mean flow and frequency spectra of the swirling jet}\label{sec:meanflow_spectra}
\begin{figure}
	\centering
  	\includegraphics[width=\textwidth]{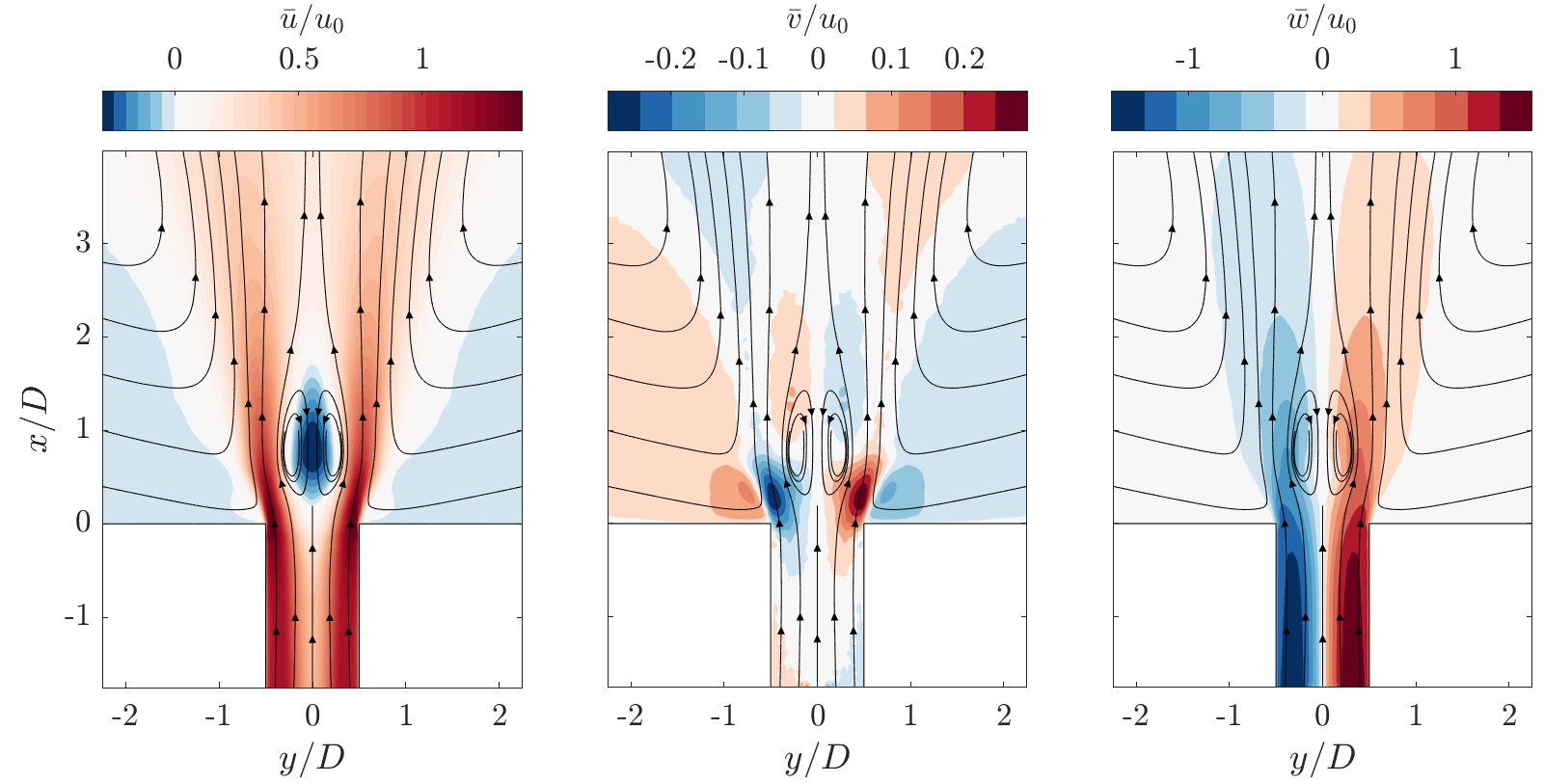}
	\caption{Normalized mean flow (left: axial, center: transverse, right: out-of-plane), $\Rey = 20 \: 000$}
    \label{fig:mean_re20k}
\end{figure}

Figure~\ref{fig:mean_re20k} shows the nondimensionalized mean flow for all three velocity components as contours overlaid with in-plane streamlines. The axial component displays the strong jet discharging into the unconfined ambience with a slow decay downstream of the duct exit. Surrounding air is entrained by the jet into the outer shear layer, which becomes particularly visible inspecting the transverse mean velocity component. The transverse component also reveals the strong initial cross-sectional spreading of the jet in radial direction when the flow exits the duct. From $x/D = 0.19$ to $1.71$ the prominent breakdown bubble can be made out in which recirculation occurs. It entails a wake-like velocity deficit downstream. The jet encloses the bubble and an inner shear layer between bubble and jet is formed. Inside the confined duct, a long extending region of axial velocity deficit in upstream direction exists as well. This significantly reduced velocity around the centerline is related to the fact that the global mode of the PVC also reaches far upstream into the duct, as will be discussed further below. To the knowledge of the authors, this important property has not been observed in any similar swirling jet setup. However, this is definitely not a necessary condition for the onset of a PVC instability. For example, in \cite{Rukes2016}, upstream of the breakdown bubble, the axial centerline velocity has indeed an excess instead of a deficit compared to the axial velocities at greater radial positions. The out-of-plane component of the mean flow shows the high rate of azimuthal rotation in positive $\theta$-direction inside the duct. Downstream of the duct exit, the out-of-plane momentum decays and is spread in radial direction due to the cross-sectional expansion, similar to the axial momentum.

\begin{figure}
	\centering
    \includegraphics[width=\textwidth]{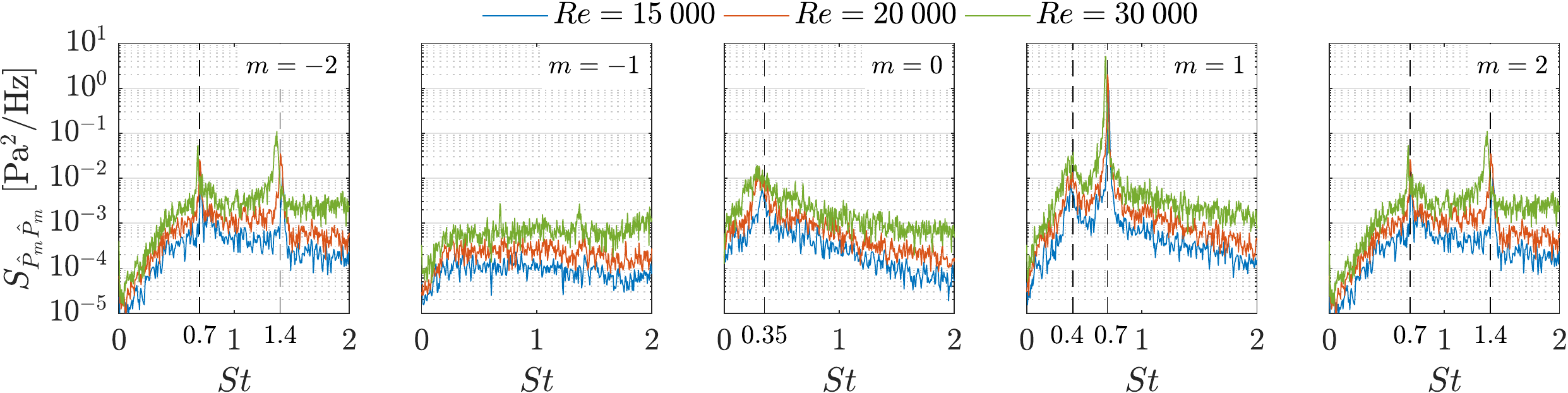}
	\caption{Power spectral density over Strouhal number $\Str$ for azimuthal modes $m = -2$ to $2$ and varied Reynolds numbers $\Rey$}
    \label{fig:psd}
\end{figure}

In the following, the PVC mode is examined in detail. In figure~\ref{fig:psd}, the power spectra are considered for all resolvable azimuthal modes of $m = -2$, $-1$, $0$, $1$ and $2$, as a function of the nondimensional frequency, the Strouhal number $\Str = f D / u_0$. The spectra are based on the time-resolved pressure measurements, as explained in section~\ref{sec:experimental_methods}. For $m = 1$, a dominant peak at $\Str \approx 0.7$ is evident for all three Reynolds numbers, corresponding to dimensional frequencies of $f = \SI{60}{\Hz}$, $\SI{78}{\Hz}$ and $\SI{115}{\Hz}$, respectively. These are the frequencies of the global, single-helical PVC mode. The first harmonic of the PVC manifests in the double-helical mode $m = 2$ at double the natural PVC frequency, $\Str \approx 1.4$, with lesser amplitude. A slight residual peak also appears at the natural PVC frequency for $m = 2$. This is due to an imperfect calibration of the pressure transducers. For the counter-rotating azimuthal mode of $m = -1$, no distinct peak exists. For $m = -2$, the same spectrum as for $m = 2$ shows up. This is due to the indeterminate phase at the limiting Nyquist wavenumber. However, it can be assumed that the rotational direction is in positive $\theta$ direction.

Another smaller peak is apparent for $m=1$. It occurs at $\Str \approx 0.4$ and is possibly related to another global mode, which has also been observed in previous works \citep{Terhaar2013b,Sieber2017}. In the present configuration this mode appears to be subcritical and it occurs in the data due to stochastic forcing. Since the peak is around two orders of magnitude lower than the PVC peak, the impact on the PVC dynamics is presumably low. At $m=0$, a small peak is visible at $\Str \approx 0.35$. It is presumably related to a weak pumping motion of the vortex breakdown bubble that has also been observed in previous studies \citep{Oberleithner2011}.\\

\subsection{The precessing vortex core and mechanisms of its formation}\label{sec:pvc_and_mechanisms}
\begin{figure}
	\begin{minipage}[t]{0.49\linewidth}
		\includegraphics[width=\textwidth]{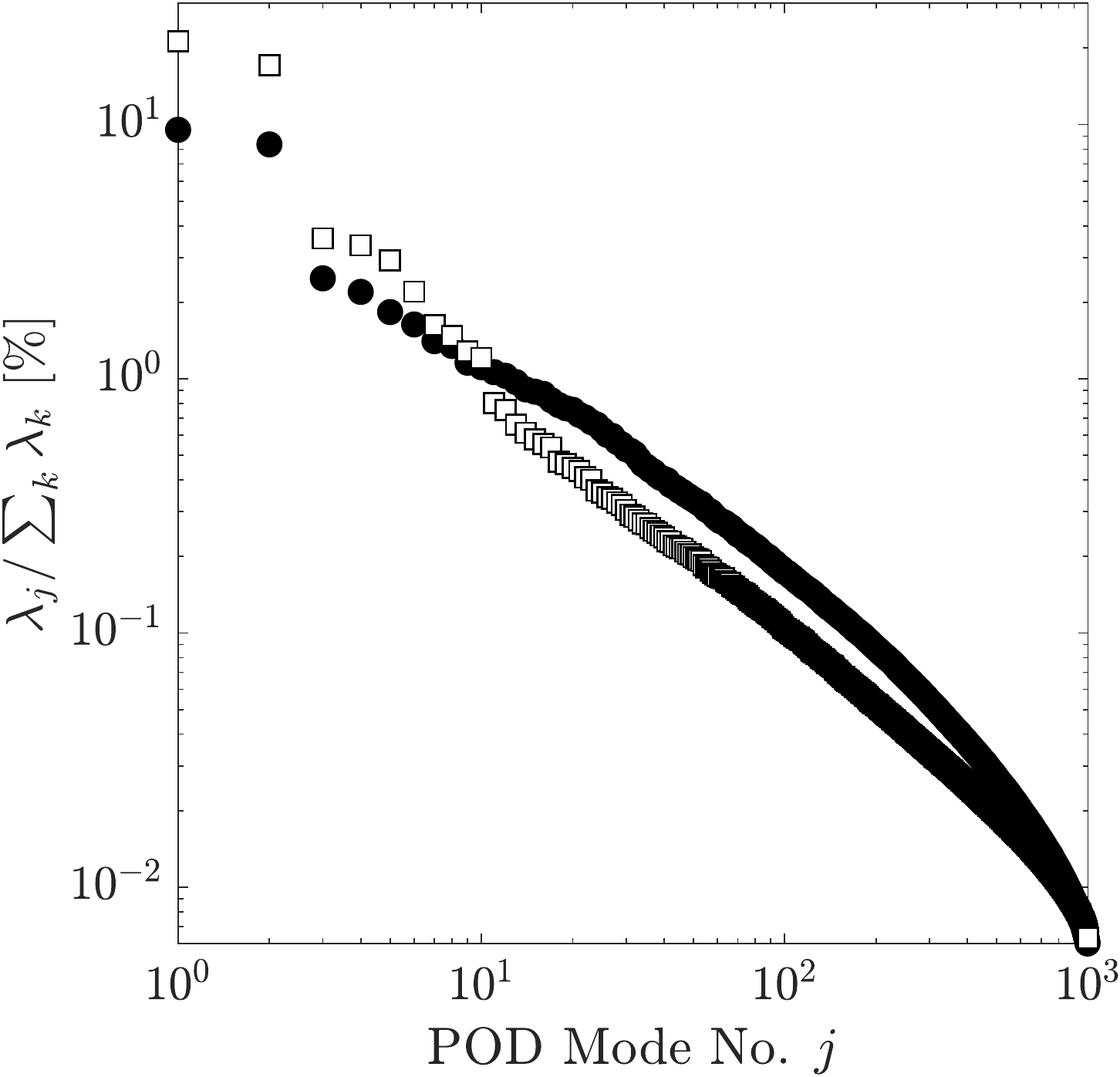}
      	\caption{POD mode energy $\lambda_j$ normalized with total turbulent kinetic energy $\sum\nolimits_k \lambda_k$, $\Rey = 20 \: 000$ ($\CIRCLE$ external domain with $x/D > 0$, $\square$ internal domain with $x/D < 0$)}\label{fig:relative_POD_mode_energy}
    \end{minipage}
    \hfill
    \begin{minipage}[t]{0.49\linewidth}
		\includegraphics[width=\textwidth]{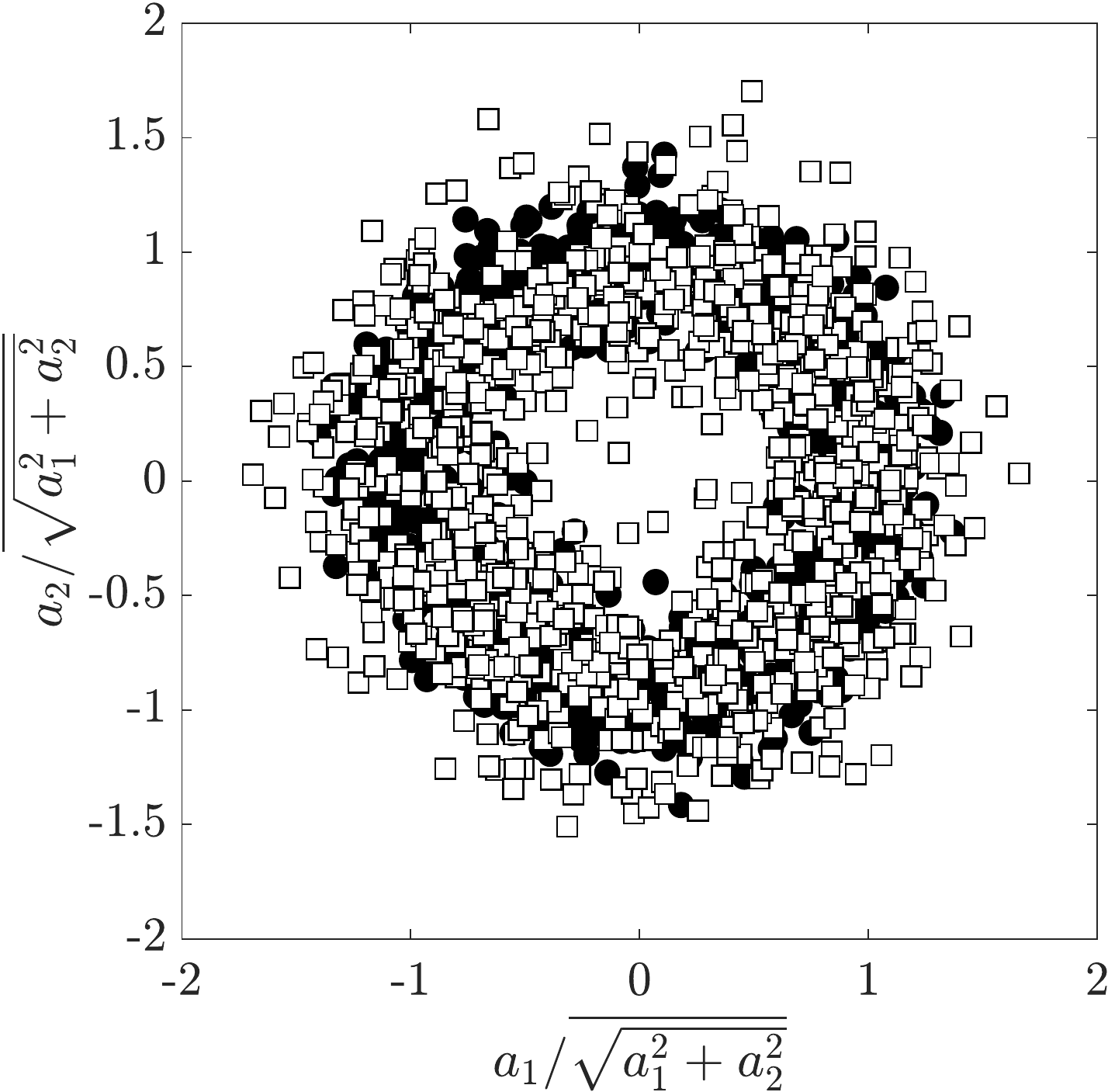}
		\caption{Phase portrait of normalized POD time coefficients for the first two most energetic POD modes, $\Rey = 20 \: 000$ ($\CIRCLE$ external domain with $x/D > 0$, $\square$ internal domain with $x/D < 0$)}\label{fig:POD_time_coefficients}
    \end{minipage}
\end{figure}

For $\Rey = 20 \: 000$, figure~\ref{fig:relative_POD_mode_energy} shows that the first POD mode pair contains much more kinetic energy than the rest of any single mode or mode pair. This applies to the coefficients inside and outside of the duct. Furthermore, the phase portrait of the time coefficients in figure~\ref{fig:POD_time_coefficients} demonstrates the harmonic nature of the first POD mode pair which is linked to the dominant peak observed in the power spectrum of figure~\ref{fig:psd}. It also explicitly shows that the harmonic nature of the PVC is clearly present inside the duct. Since the SPIV measurements are not time-resolved, the POD modes cannot be spectrally decomposed opposed to spectral POD (as e.g. proposed in \cite{Sieber2015} or \cite{Towne2018}). However, the energy optimality of the POD combined with the energy separation of the PVC relative to all other modes ensures that the leading mode pair will mainly contain spectral parts that are related to the PVC and nothing else significant. This has been demonstrated in previous studies such as in \cite{Terhaar2013b} and \cite{Tammisola2016}.

\begin{figure}
	\centering
  	\includegraphics[width=\textwidth]{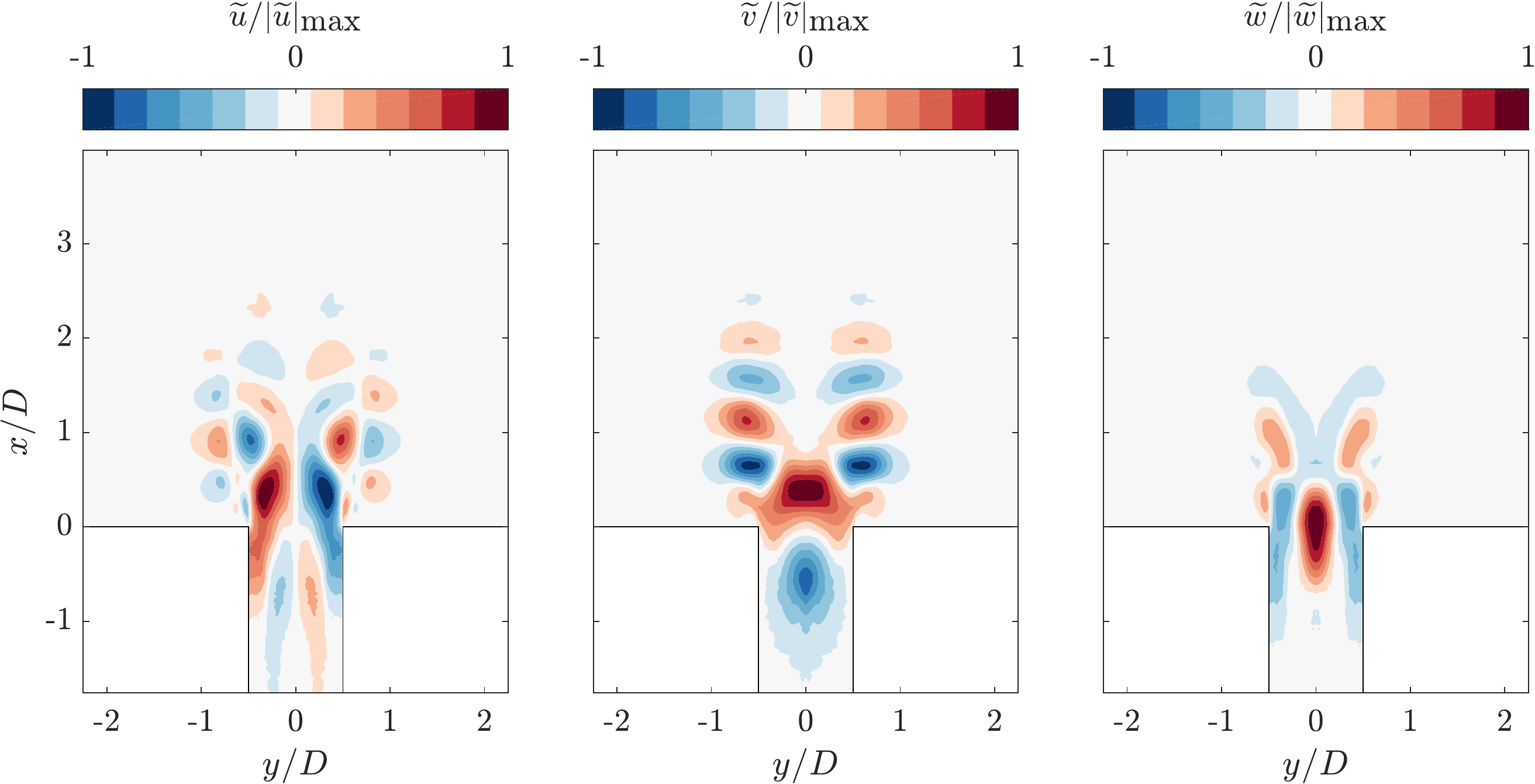}
	\caption{Normalized POD mode (left: axial, center: transverse, right: out-of-plane), $\Rey = 20 \: 000$}
    \label{fig:pod_modes_re20k}
\end{figure}

The spatial mode shape of the mode pair, according to equation~\ref{eqn:pod_superposition}, is shown in figure~\ref{fig:pod_modes_re20k} at arbitrary phase angle. The mode features nonzero velocity fluctuations in the transverse and out-of-plane component at the jet axis. From kinematic reason, this is solely possible for azimuthal modes of $m=1$ \citep{Khorrami1989}. This further establishes that the structure of this POD mode pair is associated with the dominant peak identified in the pressure spectrum at $m=1$ (figure~\ref{fig:psd}).

The PVC can be conceived as a global precession instability with which the convective Kelvin--Helmholtz instabilities synchronize. Features of both instabilities are clearly visible in the contour plots. The spatial fluctuation of the axial and transverse coherent velocity downstream of the duct exit indicates the helical shear layer vortices of the Kelvin--Helmholtz instabilities. The amplitude of the vortices decays in downstream direction after reaching a maximum around $x/D \approx 0.5$. The out-of-plane and transverse coherent velocities along the centerline describe the precession motion of the vortex core \citep{Oberleithner2011}. Most strikingly, for all three components, it is evident that the PVC mode extends far upstream into the duct, with nonzero amplitudes up to the boundary of the resolved measurement domain. The spatial amplification of the mode is already initiated here. Similar to the centerline velocity deficit of the mean flow inside the duct, to the knowledge of the authors, this important property has not been observed before.

\begin{figure}
	\centering
  	\includegraphics[width=\textwidth]{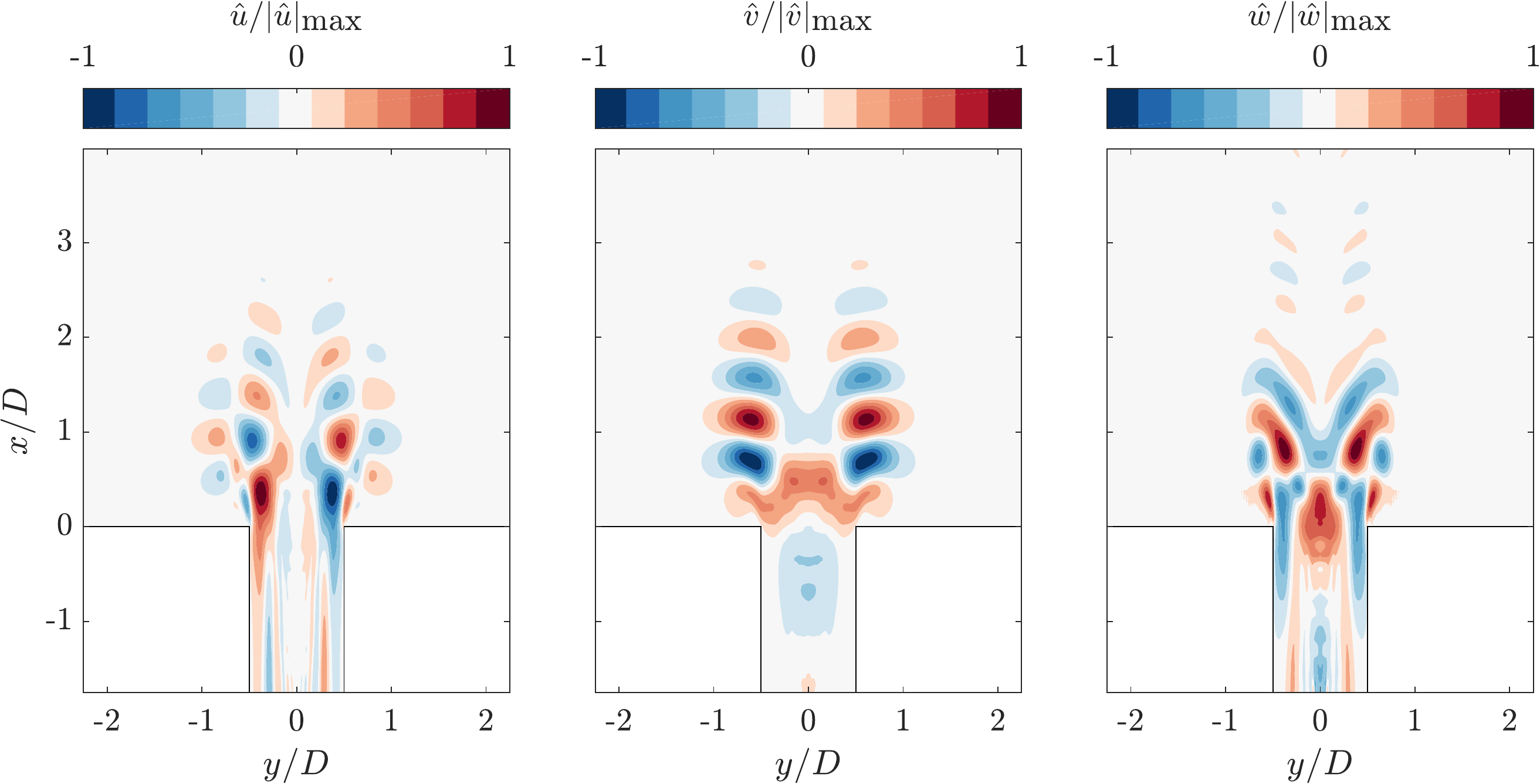}
	\caption{Normalized LSA mode (left: axial, center: transverse, right: out-of-plane), $\Rey = 20 \: 000$}
    \label{fig:lsa_modes_re20k}
\end{figure}

\begin{figure}
	\centering
  	\includegraphics[width=\textwidth]{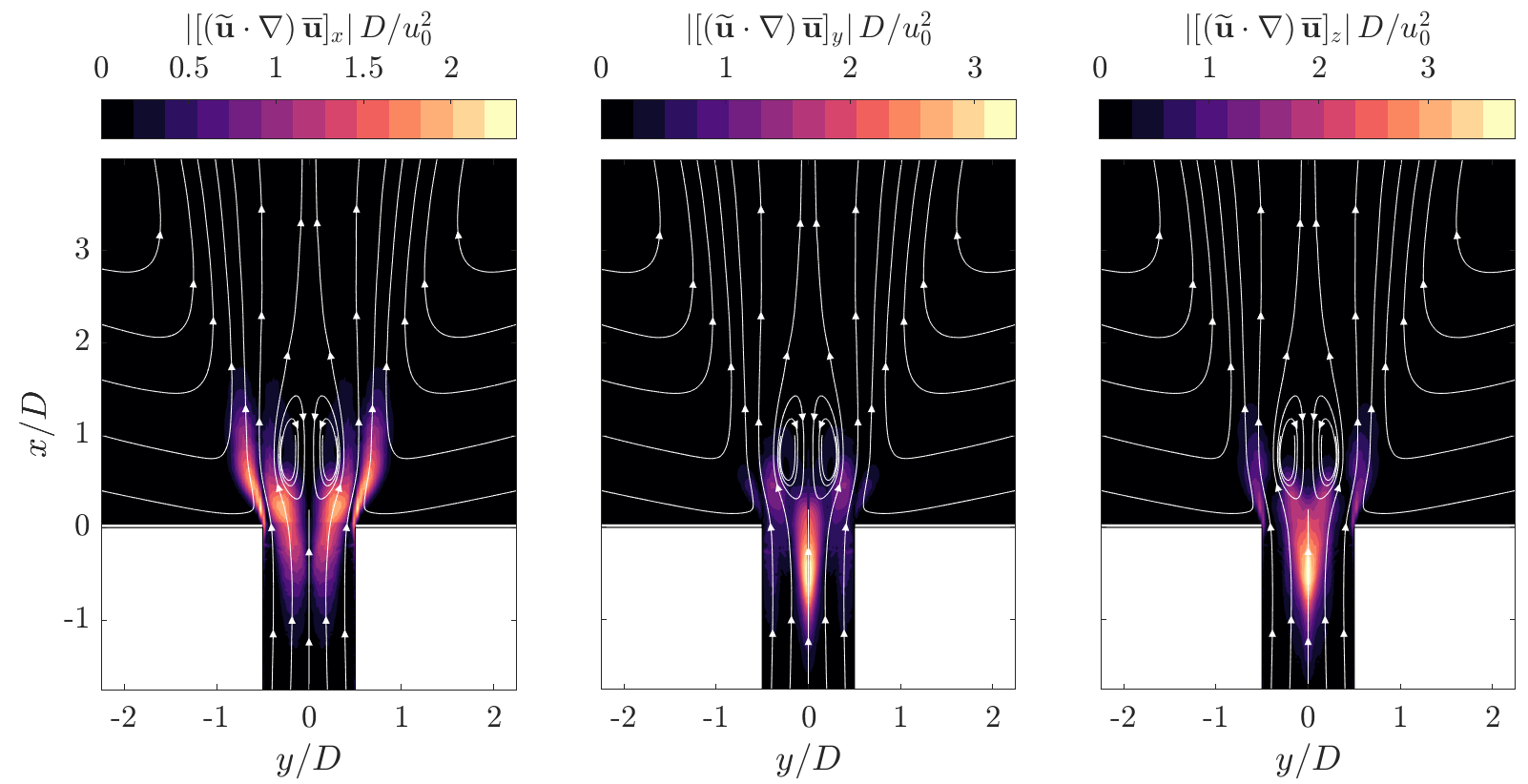}
	\caption{Coherent momentum term associated with production of perturbations (left: axial, center: transverse, right: out-of-plane), $\Rey = 20 \: 000$}
    \label{fig:production_natural_re20k}
\end{figure}

In the following, the term $(\widetilde{\vec{u}} \cdot \nabla) \, \overline{\vec{u}}$ in the Navier--Stokes equations for the coherent part (equation~\ref{eqn:coherent_nse}) is examined. This term is directly related to the production of coherent perturbations \citep{Sipp2010} and it gives insight to the mechanisms that initiate and generate the PVC. Figure~\ref{fig:production_natural_re20k} shows the magnitude of the production of coherent perturbations for each component separately. A clear spatial separation can be seen between the bulk of production of axial and the bulk of production of transverse as well as out-of-plane perturbations. The former are produced in the outer and inner shear layer of the swirling jet. While the outer shear layer is only present outside of the duct, the inner shear layer associated with the mean axial velocity deficit on the centerline is already forming inside the duct. The production of axial perturbations is attributed to the transverse velocity gradient of the mean axial velocity $\partial \overline{u} / \partial y$ which is strong in the shear layers in the vicinity of the duct exit. In contrast, the transverse velocity gradients of the mean transverse velocity $\partial \overline{v} / \partial y$ and mean out-of-plane velocity $\partial \overline{w} / \partial y$ are particularly strong on the centerline upstream of the duct exit. This gradient is responsible for the high production of both transverse and out-of-plane perturbations in this region. The significant bulk of production starts around $x/D \approx -0.8$. This region is responsible for the initiation of the precession motion. The impact of the absolute instability region of the wavemaker, which will be considered further below, apparently reaches this far upstream into the duct. Comparing the maximum values of production, it further becomes clear that the most significant contribution appears on the centerline inside the duct. This suggests that a forcing applied in these regions of high production should have the largest effect on the PVC dynamics. This will, indeed, be confirmed by the global adjoint LSA in section~\ref{sec:adjoint_and_SS}.

\subsection{Theoretical prediction of the precessing vortex core}\label{sec:pvc_direct_lsa}
\begin{figure}
    \centering
    \includegraphics[width=0.7\textwidth]{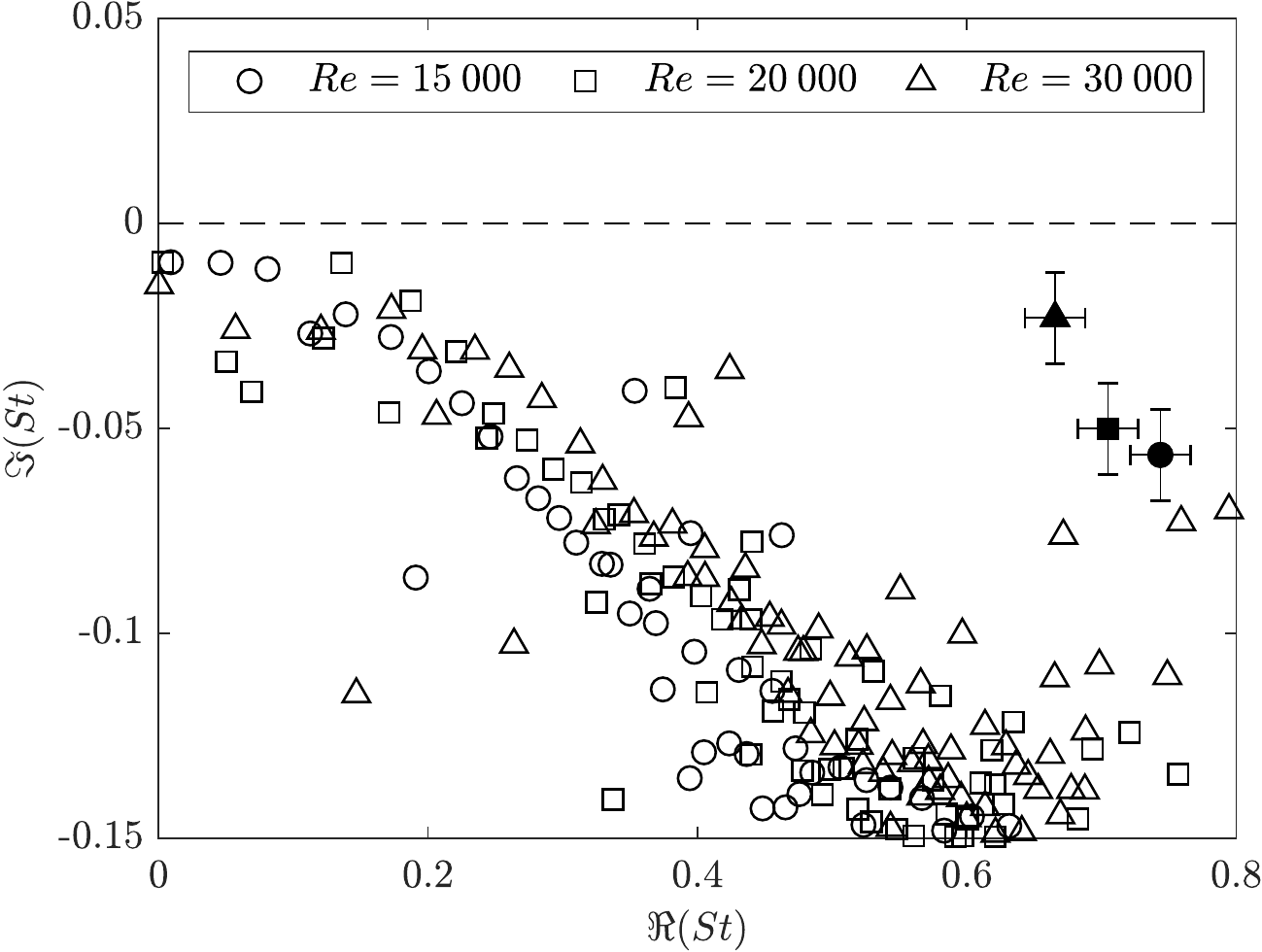}
	\caption{Eigenvalue spectrum for azimuthal wavenumber $m=1$ with real part of Strouhal number $\Real(\Str)$ denoting the non-dimensional frequency, imaginary part of Strouhal number $\Imag(\Str)$ denoting the non-dimensional growth rate; selected PVC mode (filled markers $\CIRCLE$, $\blacksquare$, $\blacktriangle$); error bars designating standard deviation of frequency and growth rate; stability limit (horizontal dashed line)}
    \label{fig:egv_spectra}
\end{figure}

The eigenvalues of the global direct LSA for all three Reynolds numbers are displayed in figure~\ref{fig:egv_spectra}. The entire spectrum exhibits stable eigenvalues only. The identified PVC modes are clustered around the experimentally measured Strouhal number of $\Str \approx 0.7$. They are discrete and do not belong to any continuous branch. With increasing mesh resolution, monotonic convergence of the selected modes is not achieved. Instead, the selected eigenvalues converge to an average value and then start to `oscillate' within a fixed limit when the mesh resolution is further increased. The lack of asymptotic convergence of the eigenvalues can be explained by the limited spatial resolution of the SPIV data. When the mesh of the global LSA is finer than the grid of the SPIV, the changes of the eigenvalues become small but seemingly `random' since the addition of grid nodes does not provide any additional information anymore. These `random' changes are attributed to discretization errors caused by the employed interpolation scheme coupled with the choice of the finite difference scheme in the LSA. Therefore, an uncertainty is associated with the eigenvalues. The error bars in figure~\ref{fig:egv_spectra} designate the standard deviation of the oscillations for both the non-dimensional frequency $\Real(\Str)$ and growth rate $\Imag(\Str)$ for the selected eigenvalues representing the PVC.

In all three cases, the growth rate is close to the stability limit but below zero. However, it is more than an order of magnitude smaller than the frequency, which means that their time scales are well separated. Furthermore, the influence of inaccuracies in the mean flow measurements and eddy viscosity estimation further affect the prediction of the true growth rate. Therefore, within uncertainty, the eigenmode of the PVC can be assumed to be marginally stable.

Other discrete modes emerge at $\Str \approx 0.4$, isolated from the continuous branches. This mode is probably related to the small additional peak that is observed in the pressure spectrum as outlined in section~\ref{sec:meanflow_spectra}. This mode may be an actual subcritical mode that is excited by stochastic forcing \citep{Farrell1993}. All other remaining modes are identified as numerical, nonphysical modes since their corresponding eigenvalues are much more sensitive with regard to discretization errors than the physical modes \citep{Sipp2010}. Furthermore, their decay rates increase with increasing mesh resolution, i.e. they become more stable.

Table~\ref{tab:LSA_eigenvalues} lists the explicit eigenvalues of the PVC obtained by the global direct LSA with their corresponding standard deviations in non-dimensional form. The LSA frequency converted to a dimensional quantity is compared to the experimentally measured frequency (as obtained by the pressure measurements, see section~\ref{sec:meanflow_spectra}). For all three cases, the relative error between LSA and experimental frequency is always below $7 \%$ within the standard deviation range.

\renewcommand{\arraystretch}{1.5}
\addtolength{\tabcolsep}{0.1cm}
\begin{table}
	\centering
  	\begin{tabular}{p{3.8cm} p{2.5cm} p{2.5cm} p{2.5cm}}
  	& $\Rey = 15 \: 000$ & $\Rey = 20 \: 000$ & $\Rey = 30 \: 000$ \\[0.15cm]
  	$\Imag(\Str_\textrm{LSA})$ & $-0.057 \pm 0.013$ & $-0.050 \pm 0.011$ & $-0.023 \pm 0.005$\\
  	$\Real(\Str_\textrm{LSA})$ & $0.744 \pm 0.024$ & $0.705 \pm 0.022$ & $0.666 \pm 0.021$\\
  	$\Real(f_\textrm{LSA})$ & $62.5 \pm \SI{2.0}{\Hz}$ & $78.9 \pm \SI{2.5}{\Hz}$ & $111.8 \pm \SI{3.5}{\Hz}$ \\
  	$f_\textrm{exp}$ & $\SI{60}{\Hz}$ & $\SI{78}{\Hz}$ & $\SI{115}{\Hz}$ \\
  	$(\Real(f_\textrm{LSA}) - f_\textrm{exp}) / f_\textrm{exp}$ & $4 \% \pm 3 \, \%$ & $1 \% \pm 3 \, \%$ & $-3 \% \pm 3 \, \%$ \\
  	\end{tabular}
    \caption{Eigenvalues of the global direct LSA and comparison of LSA frequency $\Real(f_\textrm{LSA})$ and experimental frequency $f_\textrm{exp}$ (see section.~\ref{sec:meanflow_spectra}); $\pm$ denotes standard deviation of the converged solution that starts `oscillating' around an average value when mesh resolution is further increased}
  	\label{tab:LSA_eigenvalues}
\end{table}
\addtolength{\tabcolsep}{-0.4cm}

The PVC eigenmode is shown in figure~\ref{fig:lsa_modes_re20k} for all three components at $\Rey = 20 \: 000$. In comparison to the POD mode in figure~\ref{fig:pod_modes_re20k}, a very good match exists, although there are some minor discrepancies. The amplitude of the precession motion around the centerline is underestimated inside and outside of the duct, compared to the amplitude of the Kelvin--Helmholtz vortices in the outer shear layer. However, apart from that, the overall spatial shape of the mode is captured very well including the quantitative growth and decay of the  helical Kelvin--Helmholtz vortices. Furthermore, the initiation of the precession motion inside the duct is also very well predicted by the model.

Summarizing, the relative error of the predicted frequency is very small. Furthermore, the spatial eigenmode shape is in good agreement with the experimentally obtained POD mode. The results demonstrate the validity of the global LSA in the context of turbulent swirling jets.

\subsection{Receptivity and structural sensitivity} \label{sec:adjoint_and_SS}

\begin{figure}
	\centering
  	\includegraphics[width=0.8\textwidth]{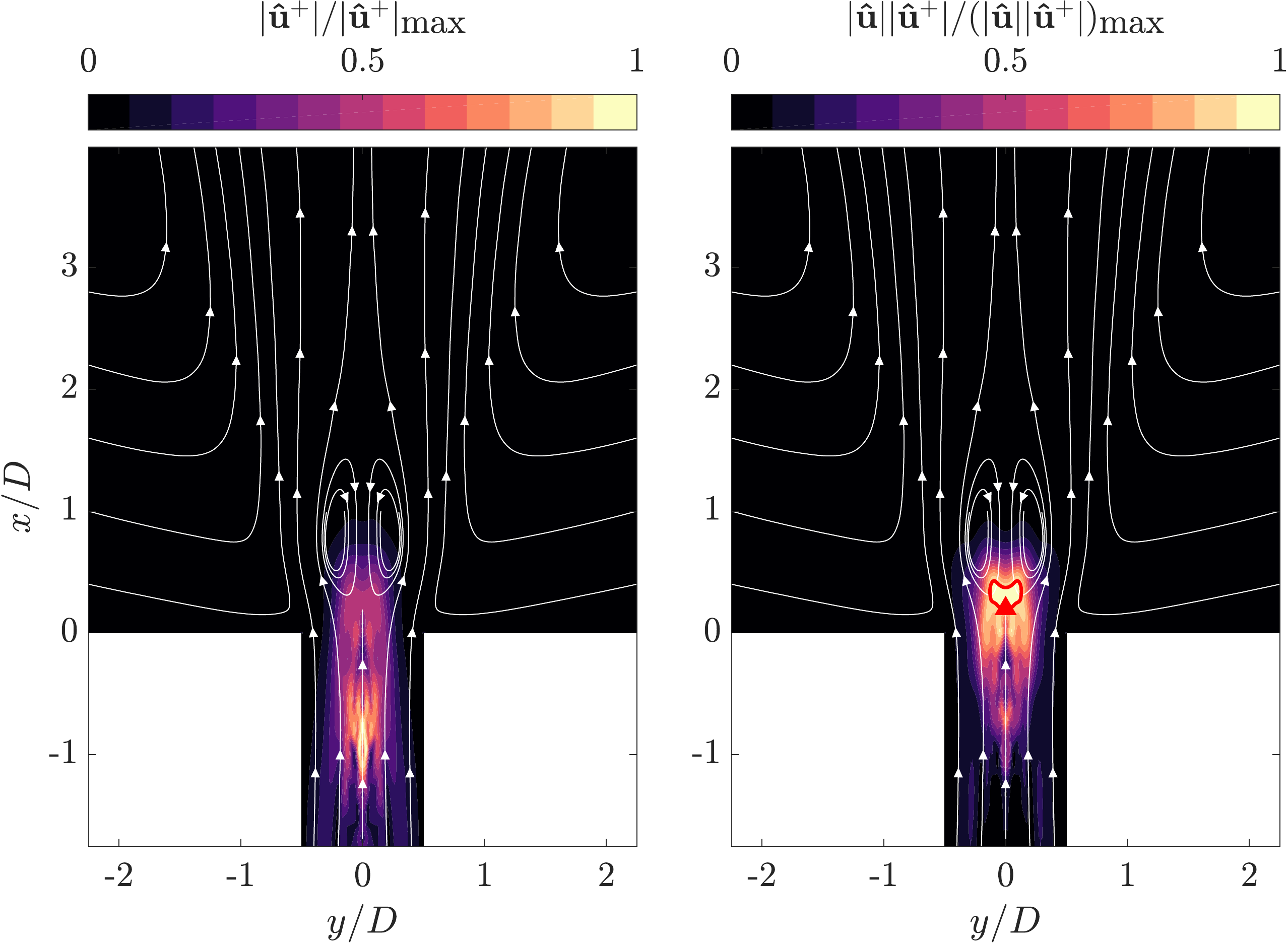}
	\caption{Normalized magnitude of the adjoint LSA mode vector indicating receptivity to open-loop forcing (left) and normalized structural sensitivity indicating sensitivity to mean flow modifications (right); the closed red line denotes regions of high structural sensitivity~$> 0.92$ ($0.21 < x/D < 0.43$); red triangle ($\blacktriangle$) denotes stagnation point of breakdown bubble ($x/D = 0.19$, $y/D = 0$), $\Rey = 20 \: 000$}
    \label{fig:adjoint_and_SS}
\end{figure}

Figure~\ref{fig:adjoint_and_SS} (left) shows the magnitude of the adjoint PVC mode. From the start of the measured domain around $x/D \approx -1.75$ downstream to $x/D \approx -1.5$, the receptivity is very low and close to zero. Hence, the PVC should only very weakly or not at all respond when actuation is applied in these regions, at least when the introduced perturbations quickly decay in downstream direction. In this case, advected residual perturbations do not survive long enough to couple with the PVC in a more receptive region. A quick decay will occur if there are no convective instabilities. Due to the observed trends, even lower receptivities can be expected upstream of the resolved measurement domain. In downstream direction, the highest receptivities are located from $x/D \approx -1$ to $-0.4$. In this region, manipulation of the PVC should be most efficient via open-loop actuation. It exceptionally well coincides with the region of maximum production of coherent perturbations (see figure~\ref{fig:production_natural_re20k}). This also explains why the receptivity is primarily accumulated close to the centerline while being almost zero close to the walls of the duct. An open-loop forcing introduced in that region would be subjected to a strong convective amplification. The largest convective growth occurs when the actuation is introduced at the start of the production around $x/D \approx -0.8$. This point overlaps with the maximum of the adjoint magnitude being reached here as well. Further downstream, the receptivity decreases slowly up to the core of the breakdown bubble. When actuation is applied further downstream, the decreasing spatial length of convective growth can be interpreted to be responsible for that trend. Thereafter, the receptivity quickly drops to zero.

The adjoint mode compares well with the adjoint mode obtained by \cite{Qadri2013} who also studied the PVC in a swirling flow for a Reynolds number two orders of magnitude lower. In their work, the amplitude of the adjoint mode reaches its maximum upstream of the breakdown bubble and concentrated around the centerline. This is also valid for the adjoint mode shown in figure~\ref{fig:adjoint_and_SS}. In their work, it is interpreted that the high receptivity is attributed to conservation of angular momentum. Any vorticity perturbation introduced in this region of high receptivity is subjected to a compression of the fluid element in axial direction and a stretching in radial and azimuthal direction (due to deceleration of the flow closely upstream of the breakdown bubble). Correspondingly, the radial and azimuthal vorticity increases while the axial vorticity decreases. This is basically the same mechanism suggested above and related to the strong production of coherent fluctuations as shown in figure~\ref{fig:production_natural_re20k}.

Based on structural perturbation theory, the structural sensitivity (figure~\ref{fig:adjoint_and_SS}, right) shows where a perturbation of the linear stability operator induces the strongest change to the mode. In other words, the structural sensitivity quantifies where in the flow the mode is most sensitive to a change of the mean flow. This is in a way equivalent to the adjoint mode that quantifies where in the flow the mode is most sensitive to an introduction of an open-loop source term.

The structural sensitivity is typically interpreted to reveal how strong the internal coupling and feedback is between receptivity of the mode and self-excitation of the mode. In the regions where the highest values are reached, the so-called `wavemaker' can be located. For the PVC, it is situated inside the breakdown bubble closely downstream of its stagnation point. The position roughly agrees with the determination of the wavemaker of similar swirling jet configurations from local LSA as well as global LSA. In \cite{Rukes2016} and \cite{Tammisola2016}, the wavemaker is located closely upstream of the stagnation point whereas in \cite{Kaiser2018} the wavemaker region is approximately situated at the stagnation point itself.

\section{Impact of open-loop control on the precessing vortex core}
In this section, it is validated whether the theoretical receptivity obtained by adjoint LSA coincides with the receptivity from the experiment. Therefore, open-loop control is applied with the goal of changing the PVC frequency, and the physical mechanisms leading to this synchronization are studied. The forcing is applied by the ZNMF actuators at different axial positions and the forcing amplitudes required for synchronization are determined via the synchronization criterion stated in equation~\ref{eqn:lockin}.

\begin{figure}
	\centering
  	\includegraphics[width=0.7\textwidth]{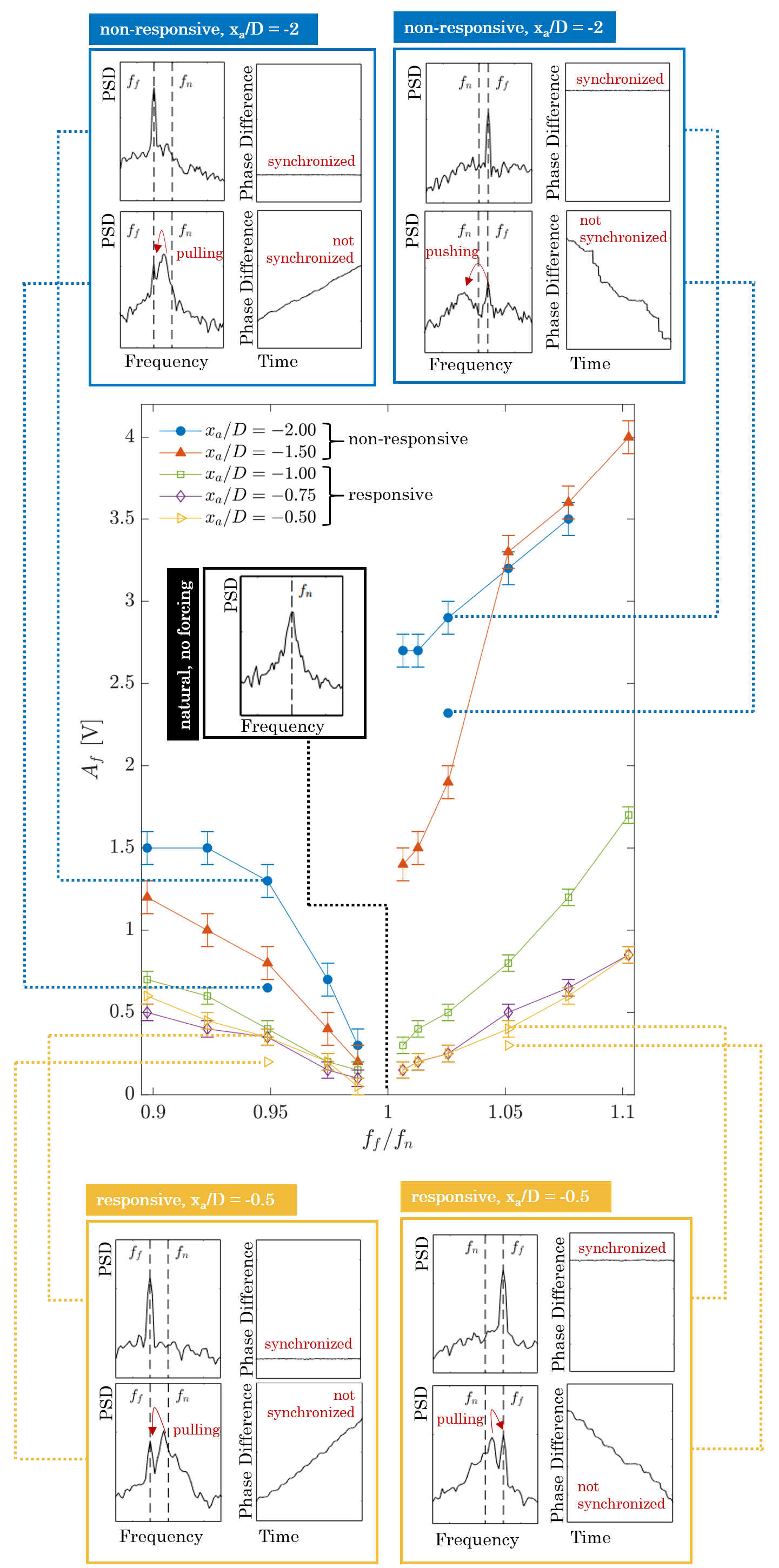}
	\caption{Forcing amplitude $A_f$ versus normalized forcing frequency $f_f/f_n$ for varied actuator positions $x_a/D$ at $\Rey = 20 \: 000$; markers with solid lines denote the minimum amplitude $A_s$ where synchronization is achieved; error bars quantify the uncertainty for determining the synchronization amplitude $A_s$, insets show PSD of the pressure measurements and phase difference for selected forcing frequencies $f_f$ and forcing amplitudes $A_f$ as indicated by the dotted lines}
    \label{fig:SFT_amplitudes}
\end{figure}

The markers with solid lines shown in figure~\ref{fig:SFT_amplitudes} depict the synchronization amplitudes $A_s$, which correspond to the minimum forcing amplitude $A_f$ where synchronization is achieved. This is shown for all five actuator positions at the Reynolds number of $\Rey = 20 \: 000$. The amplitudes are quantified  as input voltages of the actuator units. The error bars symmetrically denote the increments of forcing amplitude used in the experiments, dictating the uncertainty of the synchronization amplitudes. For the actuator positions $x_a/D = -2$ and $-1.5$, the incremental uncertainty is $\pm \SI{0.1}{\V}$. For the remaining actuator positions $x_a/D = -1$, $-0.75$ and $-0.5$, the incremental uncertainty is only half that value with $\pm \SI{0.05}{\V}$.

Generally, two synchronization regimes can be distinguished, which are denoted as the non-responsive and the responsive regime. As will be shown later, in the non-responsive regime the PVC does not respond directly to the forcing, but is only indirectly modified through modifications of the mean flow. In the responsive regime, the PVC is highly responsive, i.e. receptive, to the forcing and the PVC is easily manipulated.

The two regimes show a qualitatively different behavior when forcing is applied. The non-responsive regime can be readily identified by the strongly asymmetric behavior when forcing below the natural frequency ($f_f / f_n < 1$) and forcing above the natural frequency ($f_f / f_n > 1$) are compared. This applies to the actuator positions $x_a/D = -2$ and $-1.5$. For below-natural forcing, the synchronization amplitudes tend to go to zero when the frequency shift $\Delta f = |f_n - f_f|$ gets smaller, i.e. when the forcing frequency gets closer to the natural frequency. Inspecting the PSD of the pressure measurements for $x_a/D = -2$ reveals that the peak of the PVC is slightly pulled towards the forcing frequency prior to synchronization. Examining the phase difference over time further shows that synchronization has not been reached yet. By increasing the forcing amplitude further (i.e. the input voltage of the speakers), the peak of the PVC moves closer and closer to the forcing peak. When synchronization is reached, the two peaks are converged and the PVC stably oscillates at the forcing frequency. Then, the phase difference of both oscillators is constant over time.

In the non-responsive regime at above-natural forcing frequency, the synchronization amplitudes adopt a nonzero value when the frequency shift goes to zero. This threshold value must be exceeded even for forcing frequencies very close to the natural frequency in order to reach synchronization. The PSD prior to the synchronized state evidently shows that the residual peak of the PVC is diminished and pushed away from the forcing frequency. The phase difference indicates a non-synchronized state over time. Further increasing the forcing amplitude leads to the residual peak of the PVC being further decreased until it completely vanishes below the stochastic fluctuation level and the only dominant oscillation left in the flow is the actuation mode. Then, the phase difference indicates a synchronized state. In contrast, the responsive synchronization regime at actuator locations $x_a/D = -1$, $-0.75$ and $-0.5$ shows almost symmetric synchronization amplitudes with regard to $f_n$. Now, above-natural forcing leads to a pulling of the PVC peak towards the forcing frequency until synchronization is reached. For below-natural forcing, this pulling also occurs and is qualitatively similar to the case shown for the non-responsive regime. In both cases the synchronization amplitudes go to zero when $f_f \to f_n$. This `lock-in' behavior is typical for self-excited nonlinear oscillators that underwent a supercritical Hopf bifurcation \citep{Juniper2009,Oberleithner2011}.

All of these observed trends for $\Rey = 20 \: 000$ are very similar for the other two Reynolds numbers, except that the synchronization amplitudes are lower for $\Rey = 15 \: 000$ and higher for $\Rey = 30 \: 000$.

\subsection{Modifications to the mean flow}\label{sec:mean_flow_mod}
In this section, the changes to the mean flow are discussed which are caused by the forcing and the associated mechanisms leading to synchronization.

\begin{figure}
	\centering
  	\includegraphics[width=\textwidth]{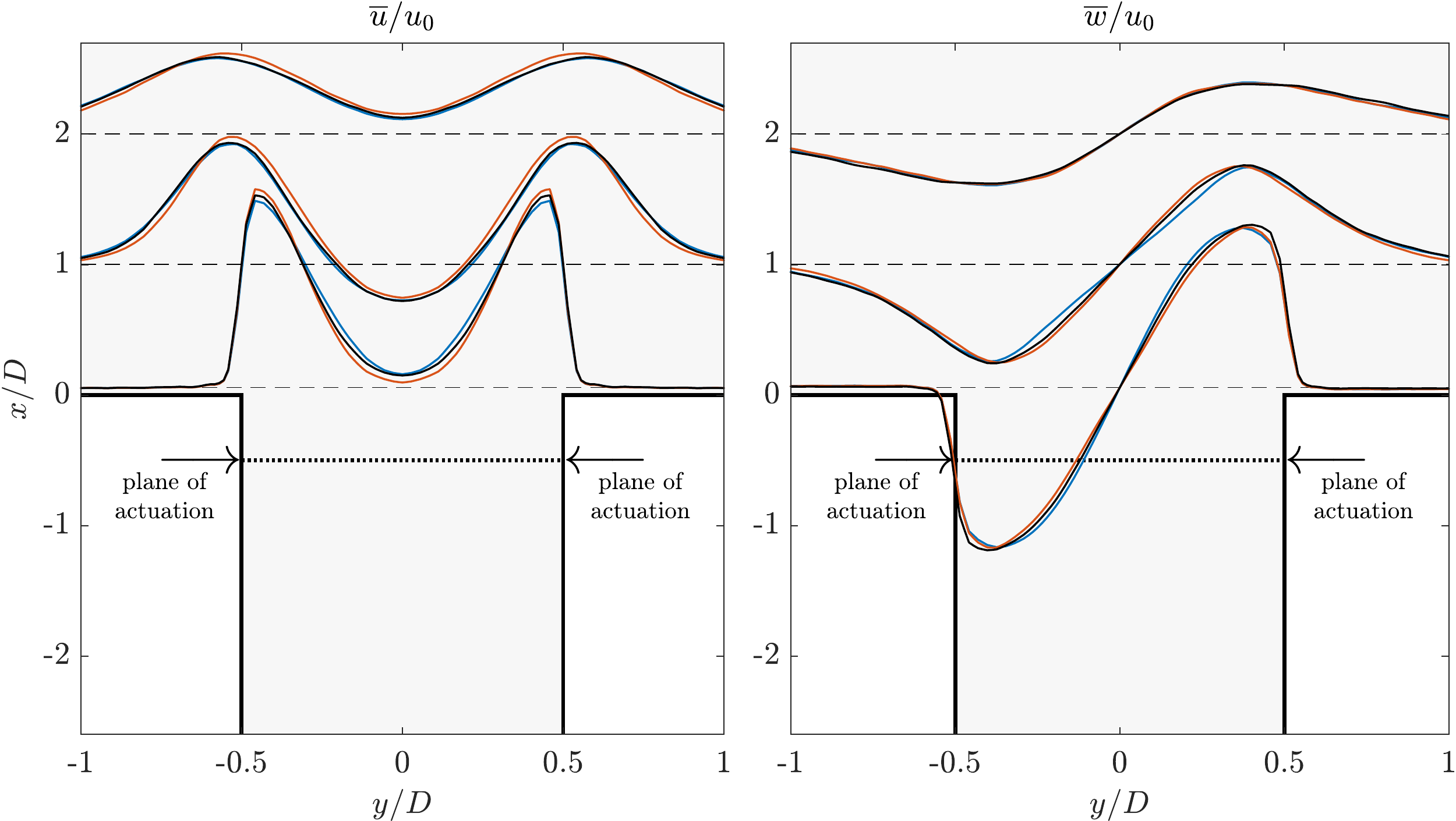}
	\caption{Mean velocity profiles at different axial positions $x/D$ for baseline (black line), below-natural forcing $f_f/f_n = 0.9$ (blue line) and above-natural forcing $f_f/f_n = 1.08$ (red line) for actuator position $x_a/D = -0.5$ when synchronization is established, for axial component (left) and out-of-plane component (right), $\Rey = 15 \: 000$}
    \label{fig:mean_axial_and_azimuthal_xaD=-0.5}
\end{figure}

Examining the mean flow fields for different forcing frequencies at the actuator position $x_a/D = -0.5$ reveals only small mean flow changes. Figure~\ref{fig:mean_axial_and_azimuthal_xaD=-0.5} indicates that the mean axial velocity on the centerline is slightly increased for below-natural forcing whereas it is slightly decreased for above-natural forcing. This is associated with small shifts of the breakdown bubble stagnation point. Below-natural forcing displaces the stagnation point downstream, above-natural forcing upstream. The mean azimuthal velocities do not significantly change. Thus, the changes to the mean flow are marginal in the responsive regime.

\begin{figure}
	\centering
  	\includegraphics[width=\textwidth]{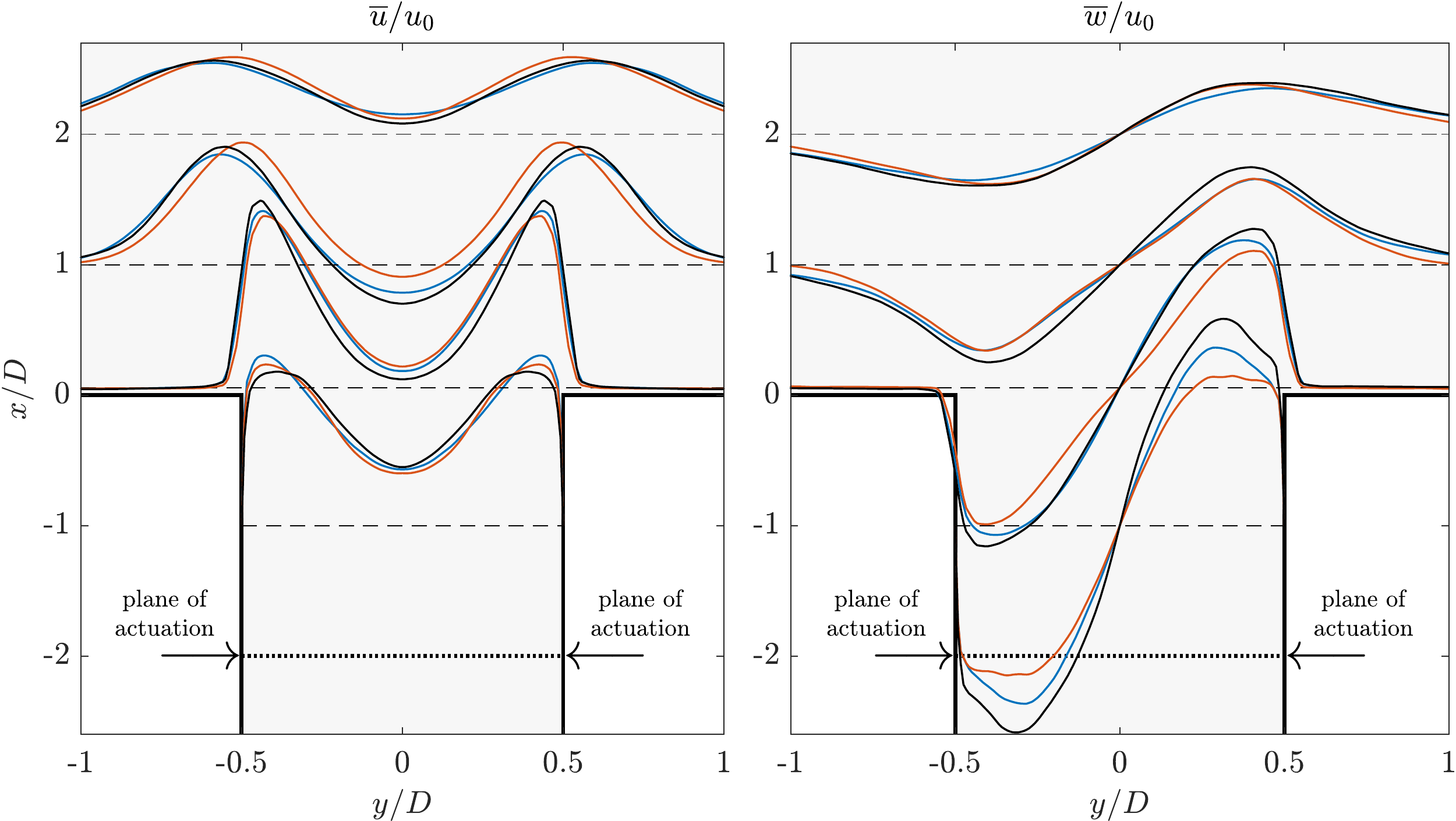}
	\caption{Mean velocity profiles at different axial positions $x/D$ for baseline (black line), below-natural forcing $f_f/f_n = 0.9$ (blue line) and above-natural forcing $f_f/f_n = 1.08$ (red line) for actuator position $x_a/D = -2$ when synchronization is established, for axial component (left) and azimuthal component (right), $\Rey = 15 \: 000$}
    \label{fig:mean_axial_and_azimuthal_xaD=-2}
\end{figure}

In contrast, figure~\ref{fig:mean_axial_and_azimuthal_xaD=-2} clearly shows that for the most upstream actuator position $x_a/D = -2$ in the non-responsive regime, the mean axial centerline velocity significantly increases at a given position for both below- and above-natural forcing. This is associated with the breakdown bubble shrinking and its stagnation point being displaced downstream. Furthermore, the maximum of the mean out-of-plane velocity is decreased in both cases.

\begin{figure}
    \centering
    \includegraphics[width=0.9\textwidth]{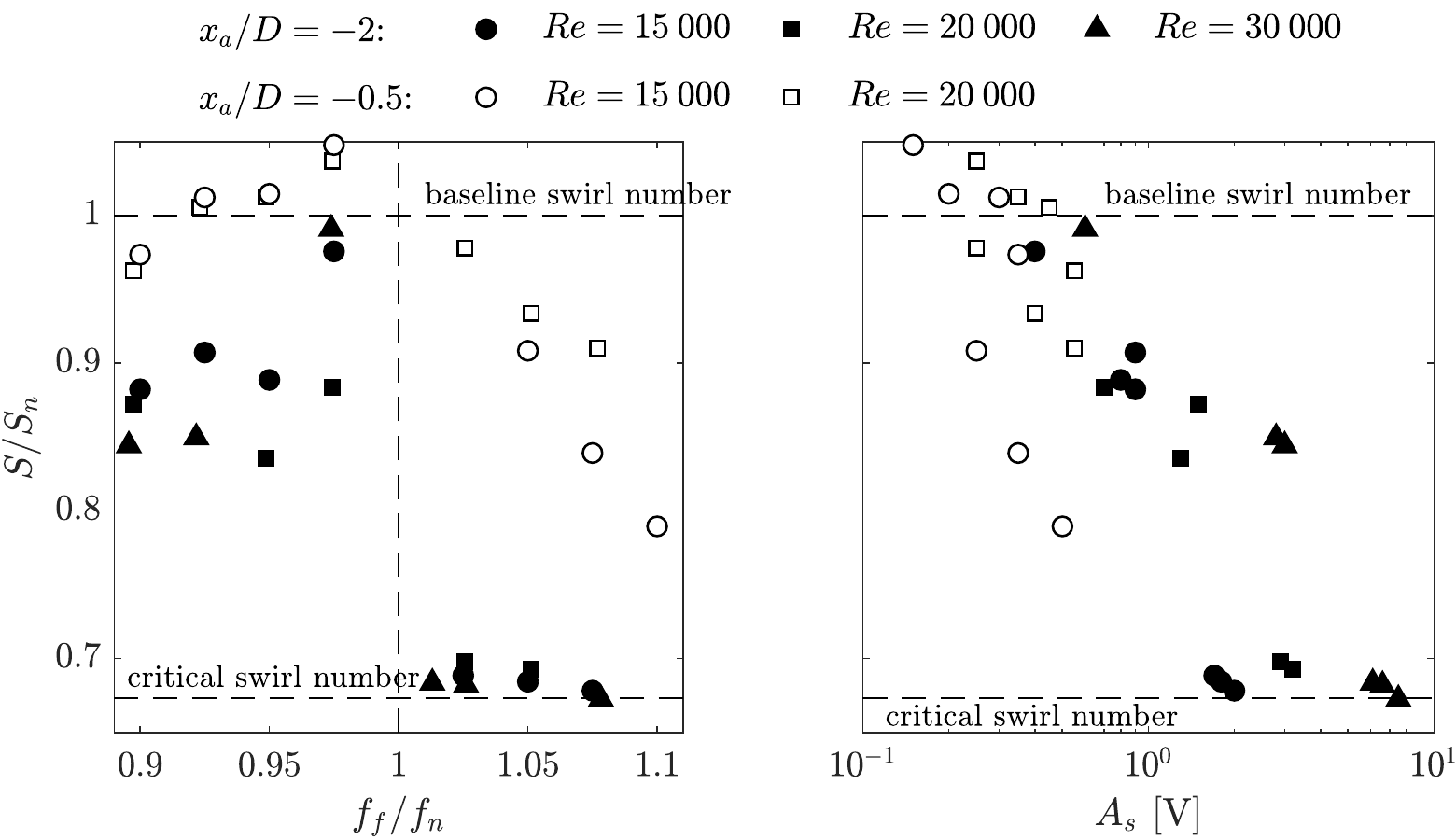}
	\caption{Swirl number modifications $S/S_n$ for synchronization as a function of normalized forcing frequency $f_f/f_n$ (left) and as a function of synchronization amplitude $A_s$ (right), actuator positions $x_a/D = -2$ and $x_a/D = -0.5$}
    \label{fig:S_changes}
\end{figure}

In figure~\ref{fig:S_changes} (left) the swirl number of the forced cases $S$ normalized on the natural swirl number of the baseline case $S_n$ is displayed as a function of normalized forcing frequency $f_f/f_n$ for actuator positions $x_a/D = -2$ (non-responsive regime) and $x_a/D = -0.5$ (responsive regime). Strikingly, the asymmetry from the tuning diagram (figure~\ref{fig:SFT_amplitudes}) is also clearly present here for $x_a/D = -2$. For below-natural forcing, the swirl number almost continuously decreases when the frequency shift $|f_n - f_f|$ is increased. On the other hand, for above-natural forcing, the swirl number is significantly reduced for every forcing frequency. The swirl number does not converge towards the natural swirl number when the frequency shift goes to zero. Furthermore, for the above-natural branch the swirl numbers are all close to the critical swirl number. The critical swirl number defines the bifurcation point that quantifies the threshold at which a steady PVC can occur for the first time when increasing the swirl number. Although the critical swirl number of the natural unmodified mean flow may not be directly comparable to the critical swirl number of the forced modified mean flow, this observation suggests that the natural PVC is suppressed in these cases.

In figure~\ref{fig:S_changes} (right, in semilog scaling), the swirl number is shown as a function of synchronization amplitude. For $x_a/D = -2$, the swirl number approximately linearly decreases with increasing synchronization amplitude for each Reynolds number, respectively. The stronger the actuation, the lower the swirl number gets. With this observation the asymmetry between below- and above-natural forcing in the non-responsive regime can be easily explained now. The asymmetry is owed to the fact that the swirl number is only a function of the synchronization amplitude in the non-responsive regime. For below-natural forcing, the swirl number reduction is in favor to the goal of reducing the frequency of the PVC. The rotation rate of the jet and, thus, the swirl number is decreased until the natural frequency associated with the swirl number matches the forcing frequency. For above-natural forcing, the swirl number reduction acts adversely to the goal of increasing the frequency of the PVC. Ultimately, a threshold of forcing amplitude needs to be exceeded in order to reach the critical swirl number and suppress the natural PVC. At this point the global oscillations are completely dictated by the actuation itself. What is then seen as a `PVC' are only Kelvin--Helmholtz vortices that still exist in the outer shear layer and that are excited at the dominating forcing frequency. Furthermore, the precession of the vortex core is artificially induced by the helical excitation of the actuator. This helical forcing substitutes the precession motion that is otherwise initiated through self-excitation in the region of absolute instability in the baseline case.

For the actuator positions of the responsive regime, it can be assumed that a strongly nonlinear interaction occurs between actuation mode and natural PVC mode due to the high receptivity at the location of actuation (see figure~\ref{fig:adjoint_and_SS}). Additionally, the trends of swirl number modification as a function of forcing frequency are clearly different to the non-responsive regime (see figure~\ref{fig:S_changes}, left). For above-natural forcing, the swirl number decreases linearly. For below-natural forcing the swirl number even increases with forcing frequencies close to the natural frequency, before linearly decreasing with a larger frequency shift. Hence, in the responsive regime there is an impact on the mean flow, even though small. Inspecting the swirl number as a function of synchronization amplitude (figure~\ref{fig:S_changes}, right), another qualitative difference to the non-responsive regime can be made out. The swirl number is not only a function of synchronization amplitude but it also depends on whether the forcing is below or above natural frequency, constituting two different branches. This suggests that there is a significant dynamical response, and not only an alteration of the mean flow. Synchronization is presumably reached by a classical phase lock-in mechanism via an inverse saddle-node bifurcation \citep{Balanov2008}. Mean flow modifications occur but are weak. They are responsible for the persisting slight asymmetry between the lower and upper branch. Again, below-natural forcing has a favoring effect while above-natural forcing has an adverse effect.

\begin{figure}
    \centering
    \includegraphics[width=0.9\textwidth]{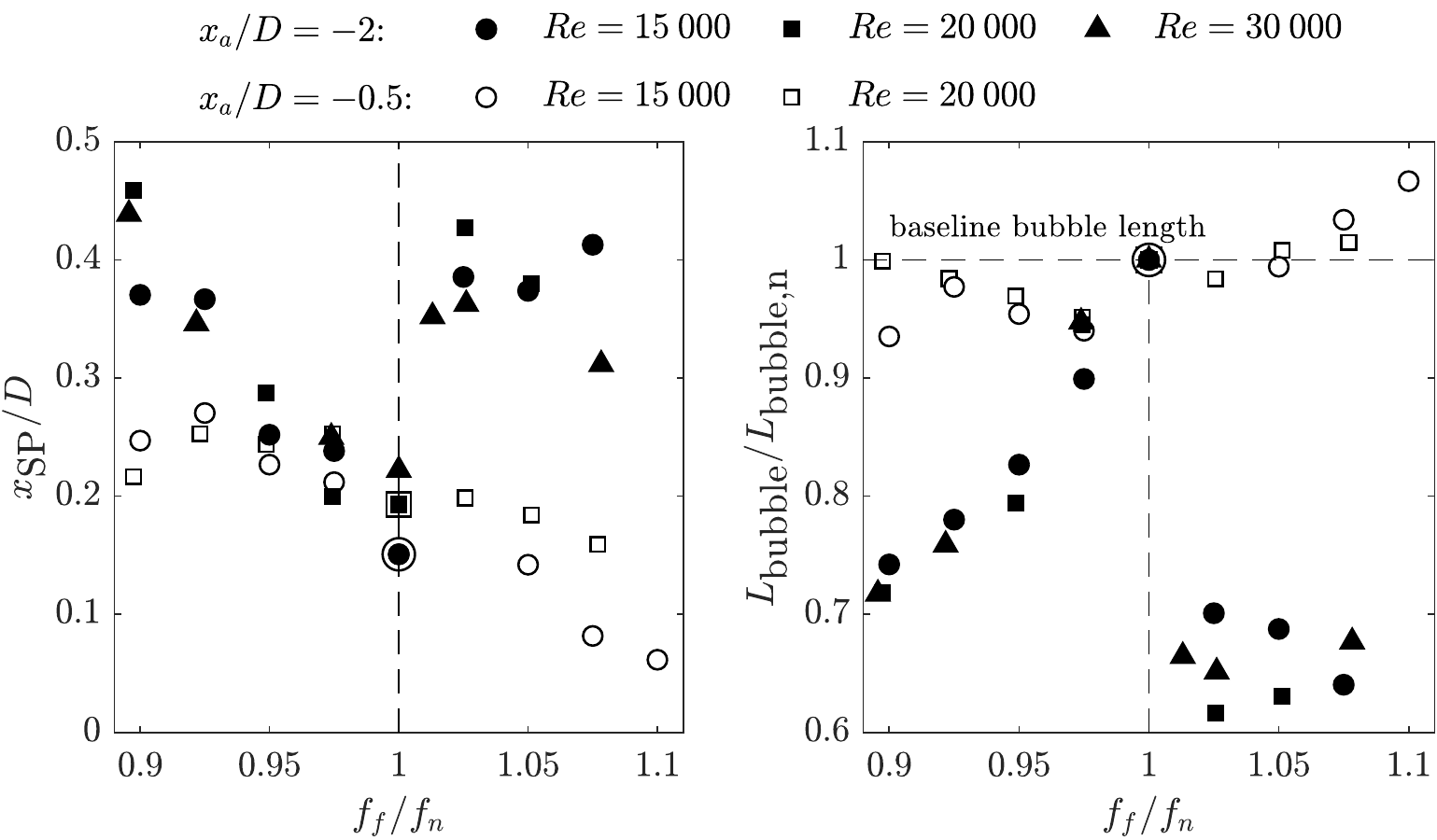}
	\caption{Changes to the stagnation point position $x_\textrm{SP}/D$ (left) and changes to the axial breakdown bubble length $L_\textrm{bubble}$ normalized with baseline bubble length $L_\textrm{bubble,n}$ as functions of normalized forcing frequency $f_f/f_n$ (right), actuator positions $x_a/D = -2$ and $x_a/D = -0.5$}
    \label{fig:x_SP_changes}
\end{figure}

Further evidence on the differences between responsive and non-responsive regime can be determined when changes to the recirculation bubble are evaluated (see figure~\ref{fig:x_SP_changes}). For the non-responsive regime, the bubble is displaced in downstream direction in every case. Since one main effect of the actuator is to dissipate rotational energy that leads to a swirl number decrease, the downstream displacement can be attributed to a delay of the vortex breakdown when the frequency shift of the forcing is increased. Accordingly, the length of the breakdown bubble in axial direction (measured from upstream stagnation point to downstream stagnation point on the centerline) is reduced. \cite{Oberleithner2012b} shows that the size of the breakdown bubble correlates with the size of the region of absolute instability. Thus, one main effect of the actuation in the non-responsive regime can be hypothesized to be a reduction of this region of absolute instability. For above-natural forcing in the non-responsive regime, the actuation is strong enough that it leads to a complete suppression of the instability. In the responsive regime, the bubble is barely displaced and the length only changes insignificantly. Therefore, the region of absolute instability likely remains untouched. This suggests that the natural mechanism is not altered, apart from the slight swirl number modifications.

The mean flow modifications occurring in the non-responsive regime clearly suggest that synchronization is not achieved by a direct interaction between actuation and PVC mode. Instead, the manipulation of the swirl number is responsible for synchronization. Either the decrease of the natural PVC frequency (below-natural forcing) or the suppression of the natural PVC (above-natural forcing) leads to synchronization. Strictly speaking, in the above-natural case, there is only a substitution of the helical PVC by the helical actuation and no synchronization takes place, since there is only one oscillator remaining in the system.

\subsection{Global linear stability of the mean flow at synchronization}
In the previous section~\ref{sec:mean_flow_mod} it was concluded that synchronization in the non-responsive regime is dictated by mean flow modifications. In order to further support this finding, a global LSA is conducted on the forced, synchronized mean flow fields. In the non-responsive regime, the global LSA should predict the correct PVC frequencies for below-natural forcing (swirl number is reduced until the modified natural frequency matches the forcing frequency) and incorrect frequencies for above-natural forcing (PVC is suppressed). Note that the LSA can only be conducted for the cases with the actuator positioned at $x_a/D = -2$ since these are the only cases where a sufficient extent of the internal duct flow was measured. This upstream domain is required for valid results of the global LSA.

\begin{figure}
    \centering
    \includegraphics[width=0.9\textwidth]{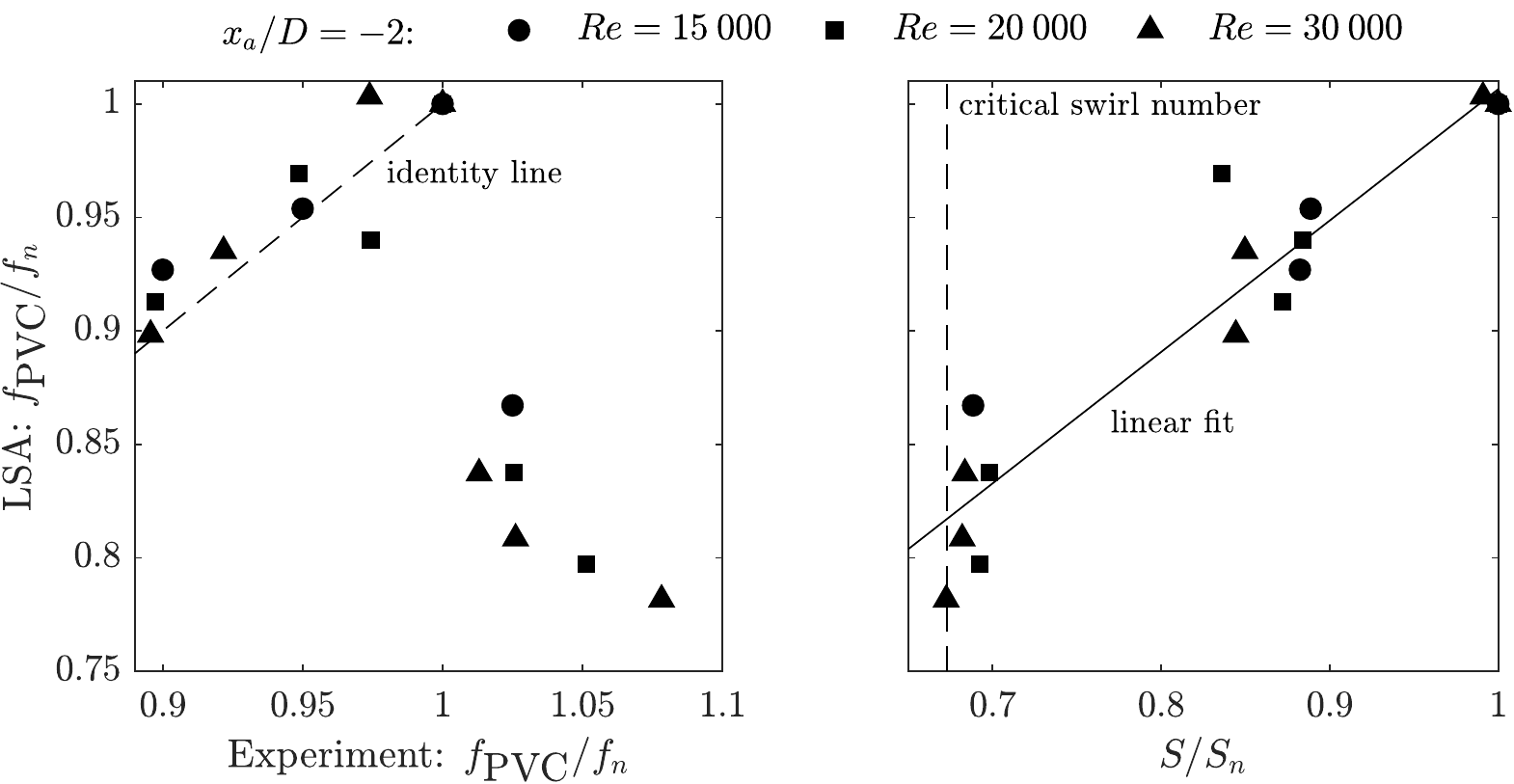}
	\caption{Global LSA on synchronized mean flows with predicted PVC frequency (obtained from LSA) compared to measured PVC frequency (obtained by experiment) (left) and compared to normalized swirl number (right), $x_a/D = -2$}
    \label{fig:LSA_sync}
\end{figure}

Figure~\ref{fig:LSA_sync} shows the results of the global LSA. On the left, the frequency obtained by the LSA is compared to the frequency obtained by the experiment. For below-natural forcing, the predicted PVC frequency approximately coincides with the measured PVC frequency. The predicted frequencies are closely gathered around the dashed identity line. Ideally, they would collapse exactly onto the identity line. However, asymptotic convergence of the solutions is not achieved and the numerical values are coupled to an uncertainty as discussed in section~\ref{sec:global_direct_lsa}. Furthermore, this effect is even more pronounced for the forced cases. As will be shown in section~\ref{sec:coherent_mod}, the coherent fluctuations are significantly large due to the forcing at the most upstream position of the SPIV resolved domain. Therefore, the setting of the inlet boundary conditions to homogeneous Neumann conditions (see section~\ref{sec:global_direct_lsa}) is questionable. For above-natural forcing, the predicted frequencies are not in line with the measured frequencies. While the measured PVC frequencies are above the natural PVC frequency, the global LSA predicts frequencies below. This demonstrates that in this case there is no actual synchronization due to swirl number modification but due to suppression. This fact is further clarified in the right plot of figure~\ref{fig:LSA_sync}. The predicted LSA frequencies for above-natural forcing are closely gathered around the critical swirl number. This shows that the PVC is suppressed. For below-natural forcing, the predicted frequencies ($S/S_n > 0.8$) increase with increasing swirl number. This is in line with the general relationship of the PVC frequency scaling proportionally to the swirl number. Both observations support that in the non-responsive regime synchronization is solely reached by mean flow and, thus, swirl number manipulation.

\subsection{Modifications to the coherent flow} \label{sec:coherent_mod}
In order to gain further insight into the response of the PVC, the modifications to the coherent fluctuations are now investigated. This will reveal whether the PVC is significantly influenced by the forcing or not.

\begin{figure}
    \centering
    \includegraphics[width=0.9\textwidth]{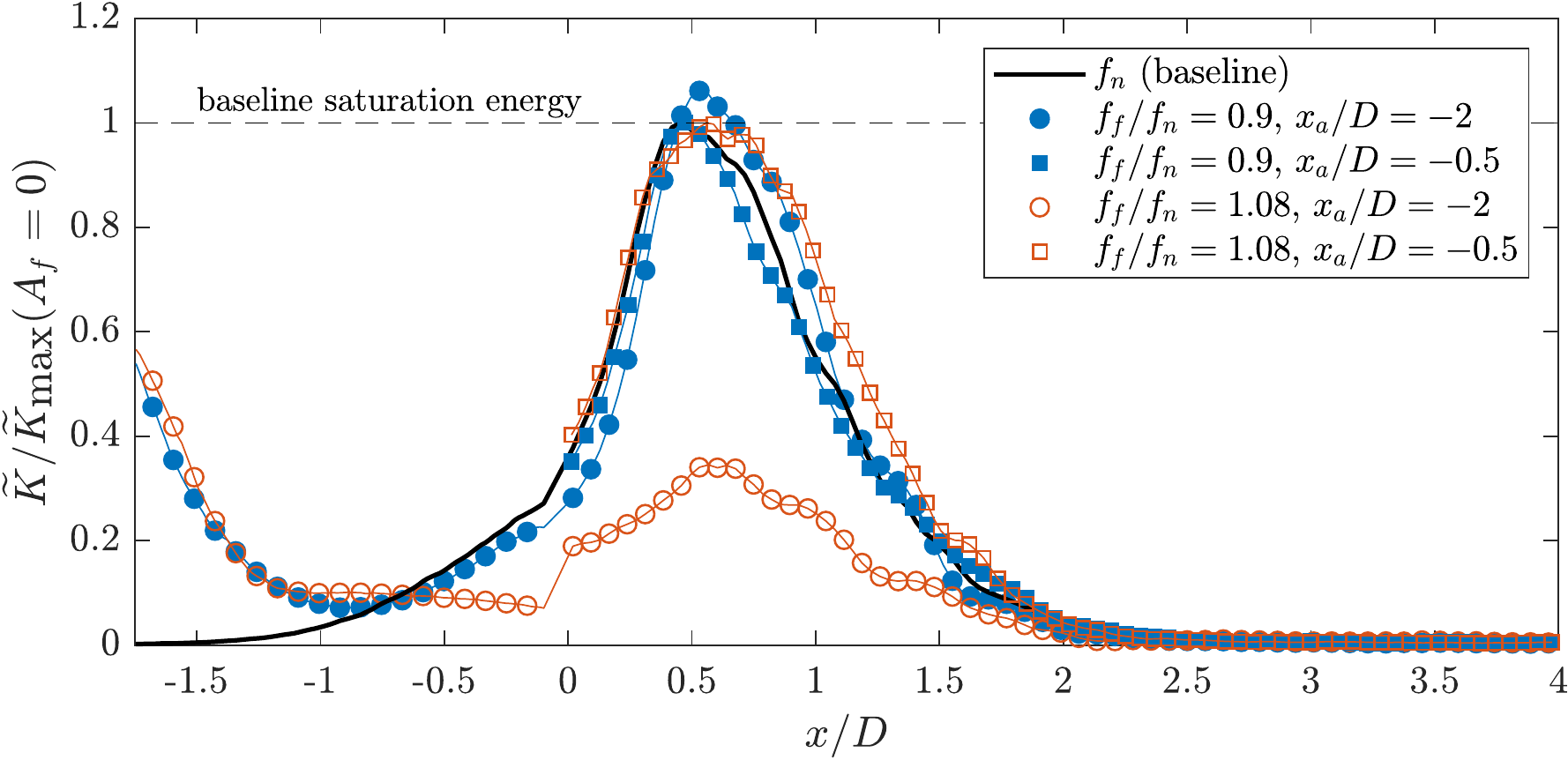}
	\caption{Total coherent kinetic energy $\widetilde{K}$ as a function of axial coordinate $x/D$ for natural PVC and synchronized PVC, actuator positions $x_a/D = -2$ and $-0.5$, $\Rey = 15 \: 000$}
    \label{fig:coherent_kinetic}
\end{figure}

At first, the total coherent kinetic energy (KE) of the PVC is used to characterize its changes caused by the forcing. It is calculated by
\begin{align}
	\widetilde{K}(x) = \frac{1}{2} \int\limits_0^{2\pi} \int\limits_0^\infty \lvert \widetilde{\vec{u}} \rvert^2 r \mathrm{d}r\mathrm{d}\theta .
\end{align}
Figure~\ref{fig:coherent_kinetic} shows the total coherent KE for the natural baseline case and for a representative below- and above-natural forced case at both $x_a/D = -2$ and $-0.5$, respectively. In the natural case, the coherent KE is almost zero at the most upstream position $x/D = -1.75$ and the amplitude is slowly convectively amplified inside the entire duct. Outside of the duct, further substantial amplification of the PVC occurs due to the generation of coherent vortices by Kelvin--Helmholtz instabilities and saturation is reached at $x/D \approx 0.5$.

When the actuator forces at $x_a/D = -2$ (non-responsive regime), the introduced perturbations increase the total coherent KE in the vicinity of the actuator. However, these perturbations decay downstream. Hence, there are no convective instabilities which are amplified at this far upstream location. For below-natural forcing, the coherent KE decays until it reaches the level of the natural case at $x/D \approx -0.8$, before the coherent KE starts to grow again. This turning point coincides with the start of major production of azimuthal perturbations observed in figure~\ref{fig:production_natural_re20k}. Downstream of this position at $x/D \approx -0.8$, the convective amplification is very similar to the natural case, however with slightly higher saturation amplitude. The similarity corroborates again that the intrinsic natural instability mechanism is barely manipulated in the case of below-natural forcing in the non-responsive regime.

For above-natural forcing, the coherent KE generated by the actuator also decays downstream until around $x/D \approx -1$. Thereafter, the energy stays at almost constant level. When the flow exits the duct, amplification of coherent fluctuations still occurs as in the natural case. However, the amplification is only related to Kelvin--Helmholtz in the outer shear layer and the saturation amplitude is significantly reduced compared to the natural saturation amplitude. This observation supports the notion that the natural PVC instability mechanism is suppressed since no convective amplification occurs inside the duct, which is in contrast to the natural and below-natural case. The precession motion is not self-excited but upheld by the continuous introduction of helical perturbations by the actuator.

For actuator position $x_a/D = -0.5$ (responsive regime), below-natural forcing does not alter the coherent energy significantly. For above-natural forcing, the axial position where saturation sets in is only slightly shifted downstream. In both cases, the saturation energy is equal to the baseline case. This supports the hypothesis that in the responsive regime the synchronization is primarily achieved by a nonlinear modal interaction between actuation and PVC. A significant alteration of the natural instability mechanism does not occur.

\begin{figure}
	\centering
  	\includegraphics[width=0.8\textwidth]{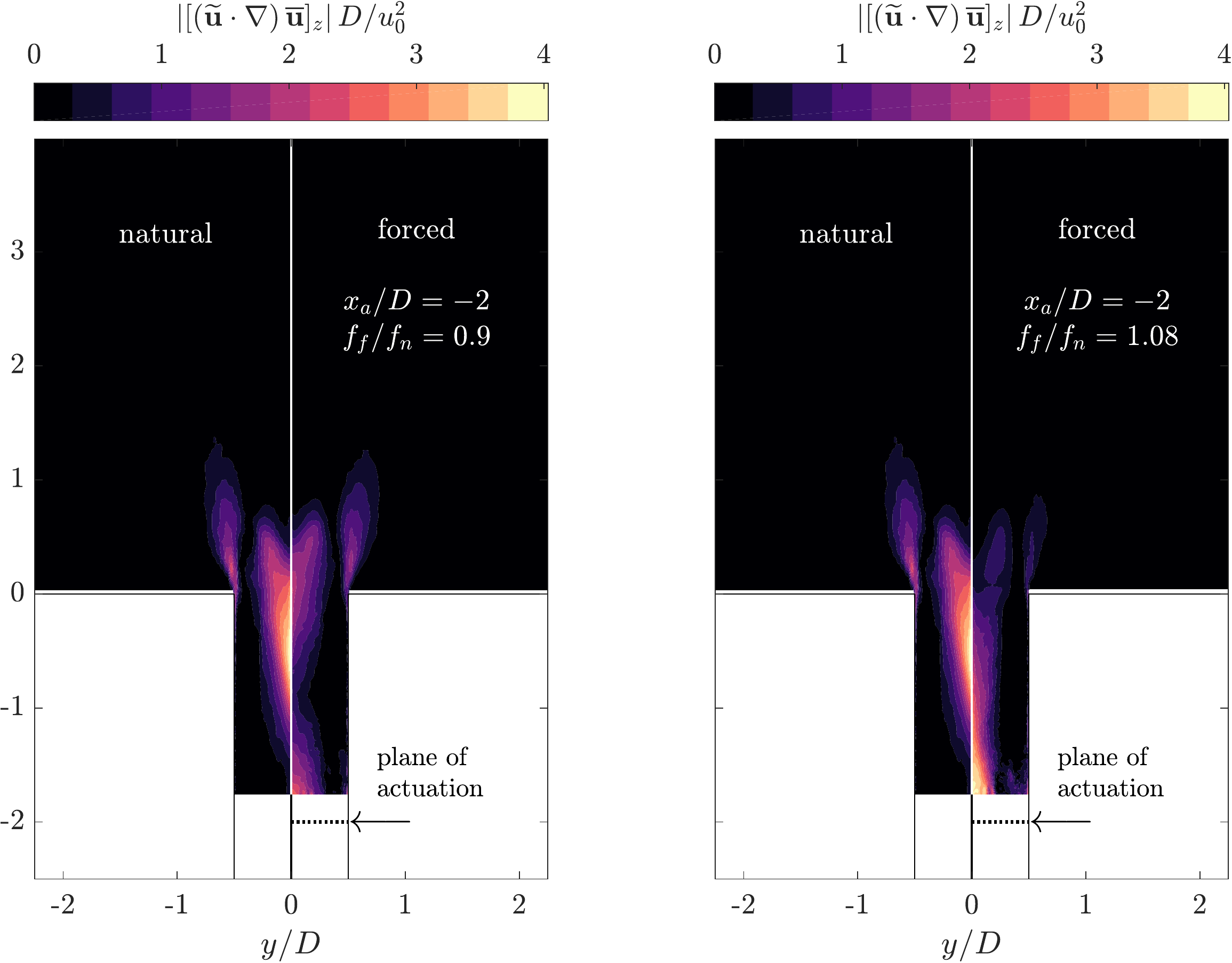}
	\caption{Impact of forcing at $x_a/D = -2$ (non-responsive regime) on coherent out-of-plane momentum term associated with production of perturbations, below-natural (left; $f_f/f_n = 0.9$) and above-natural forcing (right; $f_f/f_n = 1.08$), natural case plotted side by side to forced case, $\Rey = 15 \: 000$}
    \label{fig:production_natural_forced_xaD=-2_re15k}
\end{figure}

Now, the alterations to the production mechanism responsible for the spatial growth of the PVC mode are inspected in more detail. In figure~\ref{fig:production_natural_forced_xaD=-2_re15k} the production term is shown for below- and above-natural forcing when the actuator is placed at $x_a/D = -2$. Only the out-of-plane component is considered as representative for the precession motion (see section~\ref{sec:pvc_and_mechanisms}). For below-natural forcing, high production occurs in the vicinity of the actuator but decreases in downstream direction, reaching almost zero around $x/D \approx -1$. Downstream, the production increases again. A very similar distribution to the natural baseline case is evident with high levels of production in the inner and outer shear layer outside of the duct. Furthermore, regions of high production are found in accordance to the baseline case close to the centerline but slightly shifted downstream. Thus, the naturally acting production mechanism is not seriously altered.

In contrast, for above-natural forcing, the natural production mechanism is suppressed, as can be observed. The suppression is associated with the subcritical swirl number mentioned above. The production in the inner and outer shear layer is reduced. The maximum production occurs upstream where the actuator is situated and only decreases in magnitude downstream. In contrast to the below-natural case, there is no recovery and an increase of production in downstream direction does not happen. Hence, the bulk of production of helical perturbations originates from the actuation itself and it serves as a substitute for the self-excited perturbations that are naturally generated in the baseline case. These substitute perturbations from the actuator are then simply convected downstream (not shown), artificially sustaining the precession motion.

\begin{figure}
	\centering
  	\includegraphics[width=0.8\textwidth]{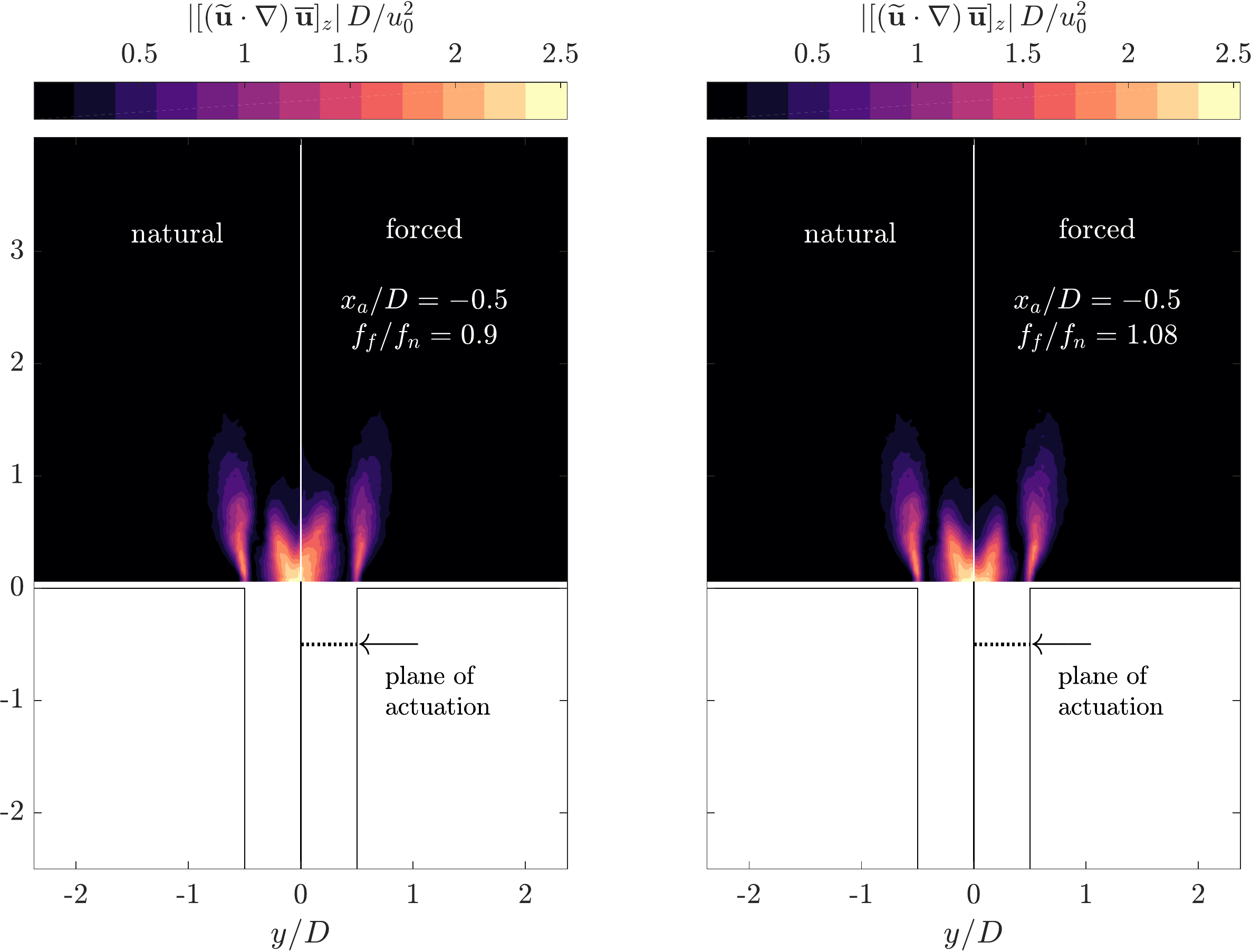}
	\caption{Impact of forcing at $x_a/D = -0.5$ (responsive regime) on coherent out-of-plane momentum term associated with production of perturbations, below-natural (left; $f_f/f_n = 0.9$) and above-natural forcing (right; $f_f/f_n = 1.08$), natural case plotted side by side to forced case, $\Rey = 15 \: 000$}
    \label{fig:production_natural_forced_xaD=-0.5_re15k}
\end{figure}

Figure~\ref{fig:production_natural_forced_xaD=-0.5_re15k} displays the production of coherent fluctuations in the responsive regime for below- and above-natural forcing, respectively, compared side by side to the natural case. In both cases, the natural mechanism is barely touched. The almost symmetric response shows that mean flow modifications play a minor role. Additionally, the fact that the production mechanism is not significantly altered suggests that synchronization cannot be attained by mean flow modification but, instead, must be reached via nonlinear interactions between actuation and PVC mode. This poses a strong contrast to the alterations of the production mechanism in the non-responsive regime.

\subsection{Comparison of experimental and theoretical receptivity}
Since the impact of open-loop control on the flow field has been investigated in detail, the experimental and theoretical receptivity can now be compared and discussed. The theoretical receptivity was defined to be proportional to the magnitude of the adjoint mode according to section~\ref{sec:receptivity_and_SS}. Consistently, the experimental receptivity is now defined to be proportional to the gradient of synchronization amplitude $A_s$ with respect to the frequency shift $\Delta f = \lvert f_n - f_f \rvert$, i.e. $\Delta A_s / \Delta f$. A larger gradient means that higher amplitudes are necessary to achieve a defined frequency shift, indicating that the flow is less receptive at the current actuator position compared to another actuator position where the gradient is lower.

The results of the synchronization diagram (figure~\ref{fig:SFT_amplitudes}) generally agree with the receptivity obtained via the adjoint PVC mode in section~\ref{sec:adjoint_and_SS}. The non-responsive regime corresponds to the adjoint PVC mode being close to zero upstream of $x/D \approx -1.5$, suggesting extremely low receptivity. Therefore, the introduced perturbations decay before the PVC responds to the open-loop actuation and only the modifications to the mean flow have a lasting effect. The mean flow modifications dictate the new frequency of the system, either by a favorable reduction of the swirl number (below-natural forcing) or by suppression of the natural PVC in conjunction with an artificial substitution of the precession motion through the actuation mode (above-natural forcing).

The responsive regime can be associated with the regions of high magnitude of the adjoint PVC mode, indicating maximum receptivity in the region of $x/D \approx -1$ to $-0.4$. Likewise, the synchronization amplitudes decrease from $x/D = -1$ to $-0.75$, suggesting an increasing receptivity of the PVC in downstream direction. Within the uncertainty range, the flow seems to have a similar receptivity for actuator positions $x_a/D = -0.75$ and $-0.5$. In these regions of high receptivity, the introduced perturbations couple with the natural production mechanism of the PVC. The perturbations are convectively amplified and overwhelm the natural dynamics, overwriting the natural frequency of the system.

The results of the synchronization studies as well as the adjoint mode quantifying the receptivity coincide with the experimental lock-in results obtained by \cite{Kuhn2016}. With a comparable setup, but the Reynolds number being an order of magnitude lower, a similar slightly asymmetrical response of the PVC is observed, requiring lower synchronization amplitudes for below-natural forcing compared to above-natural forcing. This is observed for a large number of axial positions. In the case where the actuator is traversed downstream of the breakdown bubble stagnation point, the synchronization amplitudes steeply increase. This is in line with the magnitude of the adjoint mode (figure~\ref{fig:adjoint_and_SS}) decreasing downstream of the stagnation point, indicating a declining receptivity. Likewise, for the actuator located upstream of the stagnation point within the vicinity of the nozzle exit the synchronization amplitudes stay almost constant, indicating that the receptivity of the PVC does not change much here. This can also be observed for the adjoint mode where the magnitude is approximately constant for a larger axial extent in the corresponding region.

\subsection{Synchronization mechanisms}
\begin{figure}
	\centering
  	\includegraphics[width=0.7\textwidth]{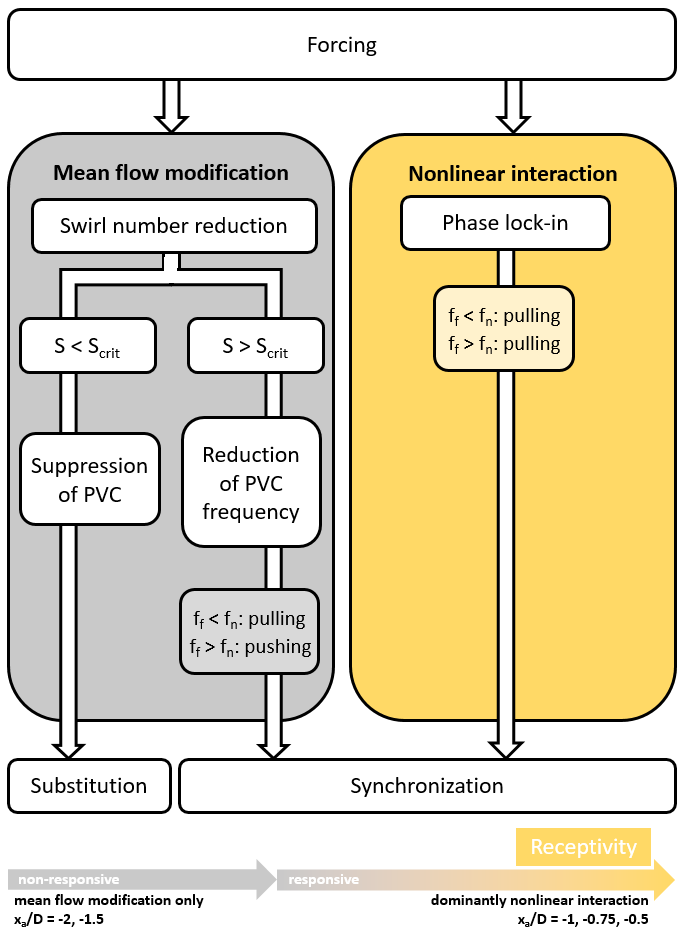}
	\caption{Schematic of physical mechanisms leading to synchronization or substitution of the PVC}
    \label{fig:flowchart_sync}
\end{figure}

In the following, the physical mechanisms that lead to synchronization in the responsive and non-responsive regime are summarized. The key mechanisms are visualized in the flowchart of figure~\ref{fig:flowchart_sync}. The forcing can have two effects on the flow: it either modifies the mean flow or it nonlinearly (directly) interacts with the PVC. The primary mean flow modification is the reduction of the swirl number by dissipation of mean azimuthal momentum. For sufficiently weak forcing, the resulting swirl number stays above the critical swirl number and the PVC frequency is decreased. This leads to frequency pulling for below-natural forcing and to frequency pushing for above-natural forcing. When the pulling becomes strong enough, synchronization can be reached. As for the pushing, synchronization can never be reached. Instead, the forcing needs to be strong enough such that the swirl number falls below the critical swirl number and the PVC is suppressed. Then, the only remaining `global mode' is artificially generated by the continuous actuation, substituting the PVC. The nonlinear interaction, on the other side, leads to a frequency pulling in any case. When the forcing is strong enough, phase lock-in sets in and synchronization is attained.

In the non-responsive regime only mean flow modifications are responsible for synchronization. This is associated with a region of virtually zero receptivity that prevents a direct nonlinear interaction. In the responsive regime, a superposition of mean flow modification and nonlinear interaction occurs. With higher receptivity of the PVC, synchronization is increasingly achieved by nonlinear interaction and mean flow modifications start to play a less important role.

\section{Conclusion}
In this work, the receptivity of a PVC in a turbulent swirling jet was characterized experimentally and theoretically. Theoretically, global adjoint LSA was conducted in order to predict locations of high and low receptivity with regard to periodic forcing. Experimentally, open-loop forcing was applied at different axial positions with the goal of finding the most efficient positions. Open questions have been tackled regarding 1) the validity of global adjoint LSA for theoretically determining the receptivity of global instabilities in the context of highly turbulent flows and 2) the physical mechanisms governing the PVC formation upstream of vortex breakdown in the confined duct, which was already suggested to play a crucial role in previous investigations.

For the experiment, a generic swirling jet setup comprising a duct of constant cross-section was considered. Open-loop forcing was realized by a ZNMF actuator at different axial positions inside the duct that helically excited the flow in order to change the frequency of the PVC, achieving synchronization between actuation and PVC oscillations. Furthermore, a global adjoint LSA was conducted on the natural, non-forced flow to obtain the theoretical receptivity in the base state.

It was demonstrated that synchronization is established via three different paths, depending on the actuator location in the duct. For the non-responsive regime, corresponding to positions with virtually zero magnitude of the adjoint modes far upstream of the duct exit, synchronization is achieved by mean flow modification in which the swirl number is reduced. For below-natural forcing, the swirl number is reduced such that the natural frequency at the modified swirl number matches the forcing frequency. For above-natural forcing, synchronization can only be achieved by the suppression of the natural PVC mechanism. Thus, the mode induced by the actuation replaces the natural PVC, artificially sustaining the precession motion of the vortex core. In the responsive regime, corresponding to positions with high magnitude of the adjoint modes closer to the duct exit, mean flow modifications are also observed. However, the modifications are small and synchronization is achieved via a classical phase lock-in mechanism in which the PVC and actuation mode directly and nonlinearly interact, resulting in both modes oscillating in phase.

The experiments showed that the flow upstream of the breakdown bubble is crucial for the formation of the PVC. Correspondingly, a finite region upstream of the duct exit was demonstrated to be very receptive to open-loop control. This high receptivity region was also predicted via global adjoint LSA. It was interpreted that the region of high receptivity correlates with the region of high production of coherent fluctuations observed on the centerline inside the duct. This region of high production is responsible for the initiation and amplification of the precession motion. Additionally, the global adjoint LSA also predicted the non-responsive region further upstream in which the adjoint mode of the PVC is virtually zero in magnitude.


The fact that the PVC is very receptive for a certain extent inside the duct may be key for future development of flow control applications. Actuators could be implemented non-intrusively into the walls of the duct upstream of the PVC. For combustors this would be particularly beneficial since the actuator could be placed sufficiently far away from the reaction zone.



This work demonstrates that global adjoint LSA is a suitable tool to estimate the receptivity of global modes in turbulent flows. The analysis correctly reveals where the mode is most receptive to open-loop forcing for flow control purposes. Although the validation method of choice (open-loop actuation) was tailored to achieve synchronization, the adjoint mode intrinsically quantifies the receptivity of the global mode to any periodic forcing. Therefore, it can be safely assumed that the receptivity also applies to closed-loop forcing where the goal is suppression through phase-opposition control. What the global adjoint LSA cannot provide is a quantitative prediction of the required synchronization amplitudes since the complete synchronization process is inherently nonlinear and the adjoint mode only provides a first-order estimate. Regardless, the adjoint mode has proved to be a valuable tool that provides satisfying estimations for the receptivity of global modes under highly turbulent conditions. \\ \\

\section*{Acknowledgments}
The support of the state of Berlin through an Elsa Neumann graduate scholarship and the support of the German Research Foundation (DFG) through grant OB 402/4-3 is gratefully acknowledged.

\section*{Declaration of interests}
The authors report no conflict of interest.

\bibliographystyle{jfm}
\bibliography{jfm-instructions,complete}

\end{document}